\documentclass[acmsmall]{acmart}
\AtBeginDocument{%
  \providecommand\BibTeX{{%
    \normalfont B\kern-0.5em{\scshape i\kern-0.25em b}\kern-0.8em\TeX}}}

\setcopyright{acmcopyright}
\copyrightyear{2025}
\acmYear{2025}
\acmDOI{XXXXXXX.XXXXXXX}

\acmJournal{JACM}
\acmVolume{37}
\acmNumber{4}
\acmArticle{111}
\acmMonth{8}

\acmPrice{15.00}
\acmISBN{978-1-4503-XXXX-X/18/06}

\usepackage{subcaption}

\usepackage{threeparttable}
\usepackage{tabularx}
\usepackage{longtable} 
\usepackage{graphicx}  
\usepackage{array} 
\usepackage{booktabs} % For \toprule, \midrule, \bottomrule
\usepackage{tcolorbox}
\usepackage{multirow}

\usepackage{amssymb}
\usepackage{tikz}
\usetikzlibrary{shapes.geometric, arrows, positioning} 
\tikzstyle{block} = [rectangle, draw, text centered, rounded corners, minimum height=2em]
\tikzstyle{arrow} = [thick,->,>=stealth]
\tikzstyle{cloud} = [ellipse, draw, text centered, minimum height=2em]
\usetikzlibrary{arrows.meta,positioning}
\usetikzlibrary{shapes.callouts, positioning}
\usepackage{fontawesome5}
\usetikzlibrary{shapes.callouts}
\usepackage{fontawesome5}
\begin{document}

\newcommand{\subsubsubsection}[1]{\textbf{#1.}}

\title[Pinning Inappropriate Comments as a Moderation Strategy]{The Pin of Shame: Examining Content Creators' Adoption of Pinning Inappropriate Comments as a Moderation Strategy}
%Your Comment Got the Pin of Shame! Exploring Comment Pinning as a Tool for Content Moderation.

\author{Yunhee Shim}
\email{yunhee.shim@rutgers.edu}
\affiliation{%
  \institution{Rutgers University}
  \city{New Brunswick, NJ}
  \country{USA}
  }

\author{Shagun Jhaver}
\email{shagun.jhaver@rutgers.edu}
\affiliation{%
  \institution{Rutgers University}
  \city{New Brunswick, NJ}
  \country{USA}
  }

\renewcommand{\shortauthors}{Shim and Jhaver}

\begin{abstract}
Many social media platforms allow content creators to \textit{pin} user comments 
%in response to their content. 
to their posts.
Once pinned, a comment remains fixed at the top of the comments section, regardless of subsequent activity or the selected sorting order. 
The ``Pin of Shame'' refers to an innovative re-purposing of this feature, where creators intentionally pin norm-violating comments to spotlight them and prompt shaming responses from their audiences.
This study explores how creators adopt this emerging moderation tactic, examining their motivations, its outcomes, and how it compares—procedurally and in effect—to other content moderation strategies. 
Interviewing 20 content creators who had pinned negative comments on their posts, we find that the Pin of Shame is used to punish and educate inappropriate commenters, elicit emotional accountability, provoke audience negotiation of community norms, and support creators’ impression management goals. 
Our findings shed light on the benefits, precarities, and risks of using public shaming as a justice practice in platform governance. 
We contribute to HCI research by informing the design of user-centered tools for addressing content-based harm.
% by advancing our understanding of user-driven content moderation.
% It highlights how users’ behavioral creativity challenges traditional moderation frameworks and offers design implications such as emotion-sensitive engagement metrics and increased transparency in the sanctioning process. 
% Ultimately, this work provides insights into how moderation systems can be designed to incorporate transparency and promote healthier online discourse.

%Abstract of 100-150 words
\end{abstract}




\maketitle
% Paper structure

\section{Introduction}
Jayan is a content creator who regularly shares fitness content on Instagram, where he has over 29,000 followers. 
On March 24, 2025, he experienced an unexpected viral moment when one of his videos reached 140,000 views and attracted more than 12,000 comments. 
While skimming through the comments section, he came across a particularly offensive comment: ``I wake up every day happy knowing I’m not Indian \includegraphics[height=0.8em]{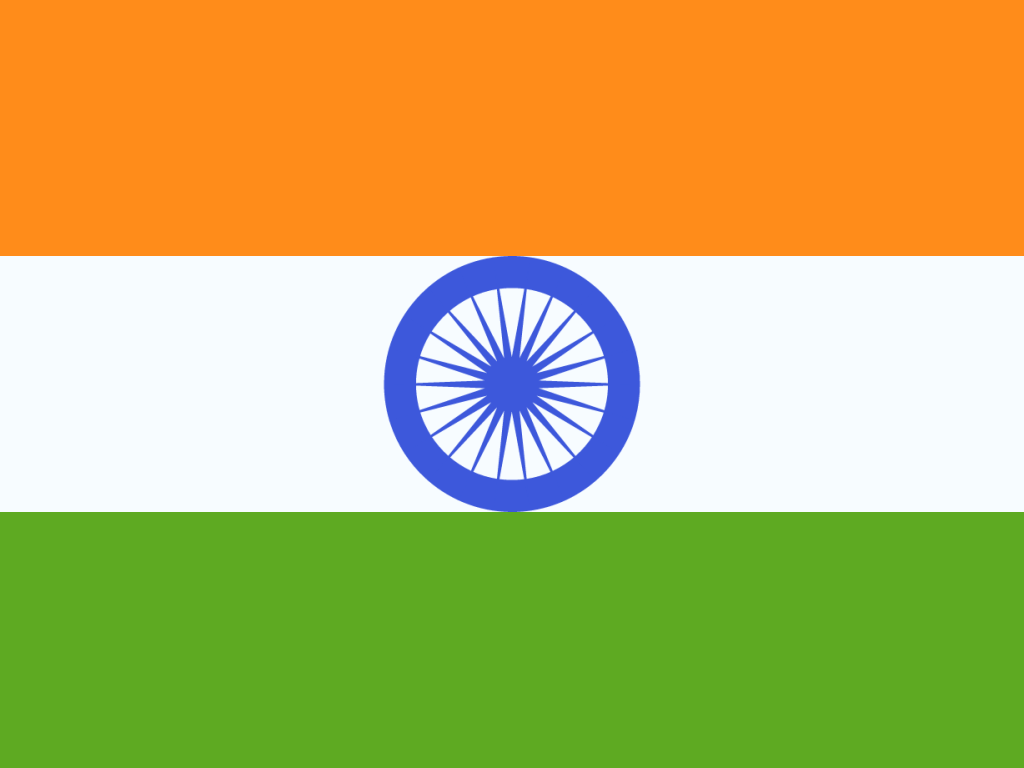} \includegraphics[height=0.9em]{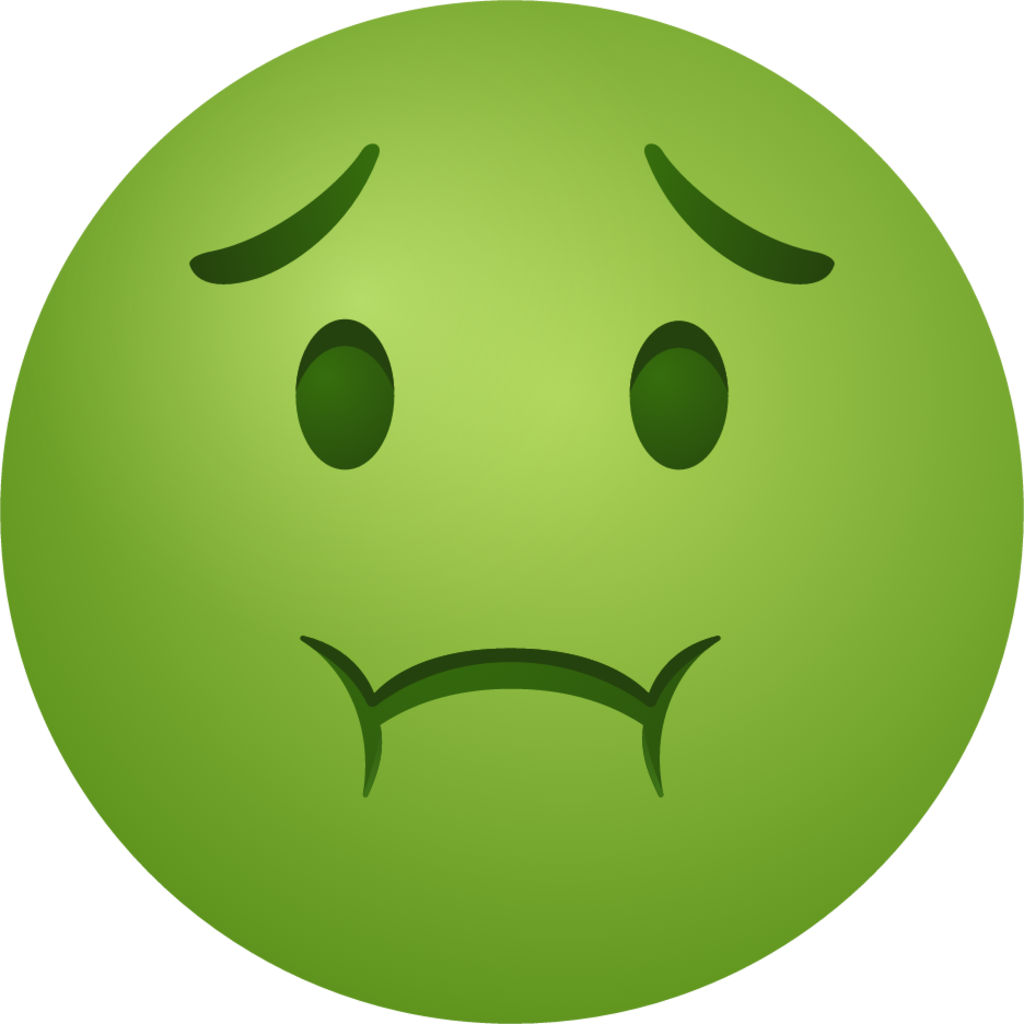} \includegraphics[height=0.9em]{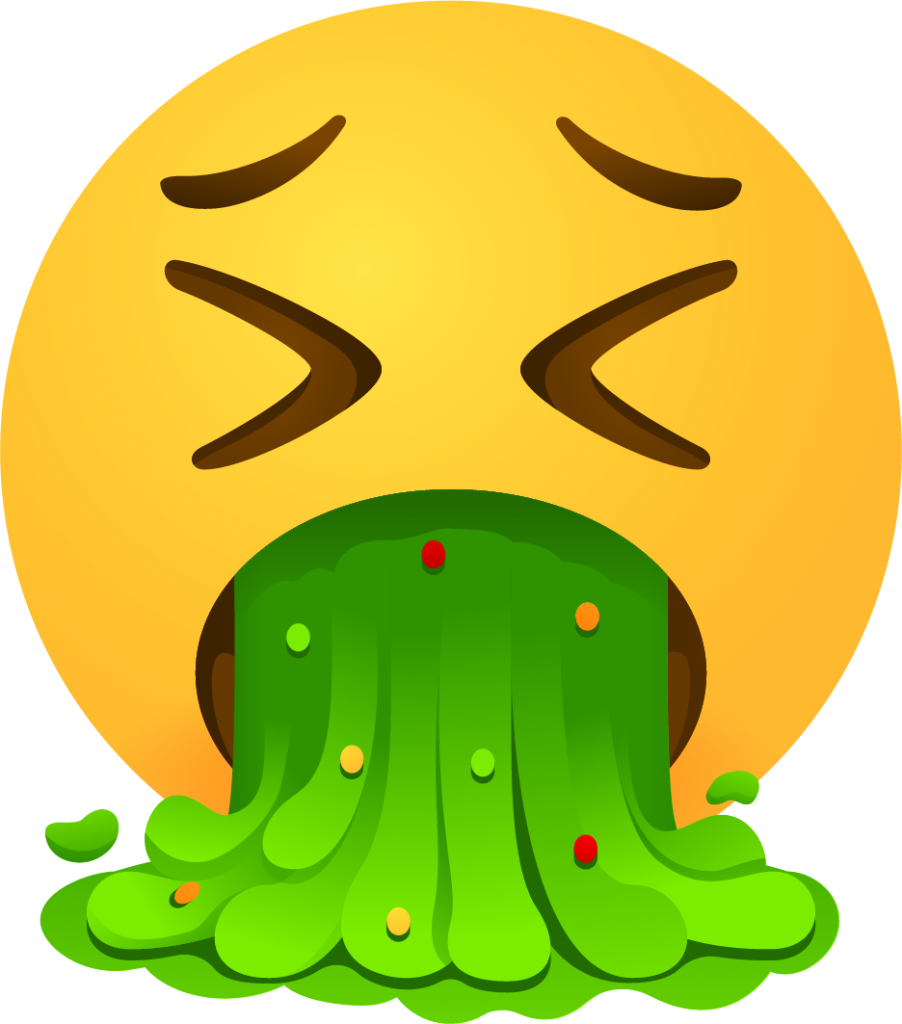},''.\footnote{We received permission from Jayan to describe this incident in the current paper.}
Although Jayan had witnessed similar personal attack comments on his previous posts, this instance felt especially insulting---it attacked not just him, but also his heritage. 
In response, he pinned the comment to the top of the comments section---an action now common enough that many Instagrammers recognize it and colloquially refer to it as the \textit{`Pin of shame'}. 
This act prompted his followers to support him via posting replies that shamed the pinned commenter, resulting in over 100 follow-up responses directed at the negative commenter.
This support validated Jayan's outrage at the offensive comment and helped ease his distress.
While the shamed commenter did not respond to this backlash, they silently deleted their comment after some time, which removed the entire comment thread (i.e., the offensive comment and all replies to it) from the comments section.

Offensive comments such as the one Jayan received are a frequent occurrence for creators (i.e., actively posting social media users with a large number of followers) on social media platforms~\cite{jhaver2022designing,salehi2023sustained,lyu2024upload}.
%Online platform users may encounter hostile and malicious behavior, such as online hate speech and harassment, which sometimes escalate as a significant social problem ~\cite{lowry2016adults,jhaver2018online}. 
%Many creators are frequently exposed to hate speech and harassment. 
For example, ~\citet{thomas2022s} surveyed 135 creators and found that  95\% of them had experienced an incident of hate and harassment at least once.
Prior HCI research has characterized the viewing of such comments as \textit{content-based harm} and noted how it can lead to anxiety and mental health risks~\cite{jhaver2022designing}.
Researchers have also examined how creators respond to such incidents and evaluated the utility of \textit{creator-led content moderation tools} that platforms offer to counteract offenders, such as blocking~\cite{geiger2016bot,john2024classification}, reporting~\cite{crawford2016flag,zhang2023cleaning}, and word filters~\cite{jhaver2022designing}.
Findings from this research show that these tools present several challenges for the creators, including the cognitive labor required to use them, lack of control and transparency about their operations, and novel privacy concerns that emerge through their deployment~\cite{jhaver2022designing,zhang2023cleaning,jhaver2023personalizing}.

%negative commenter as seen in the case of Jayan above.
% We observed that the types of norm violations of the pinned comments can vary, including identity-based group discrimination, mocking the content creators, and personal attacks (see Figure~\ref{fig:example}). 
% Additionally, the way of deploying the Pin of Shame behaviors also varied: Creators sometimes pin the comment and leave it without any interaction (left), signal the inappropriate commenter with remarks of `Pin of Shame' (middle), or clap back with witty or sarcastic reply (right).

\begin{figure}
    \centering
    \fbox{\includegraphics[width = 4.1cm]{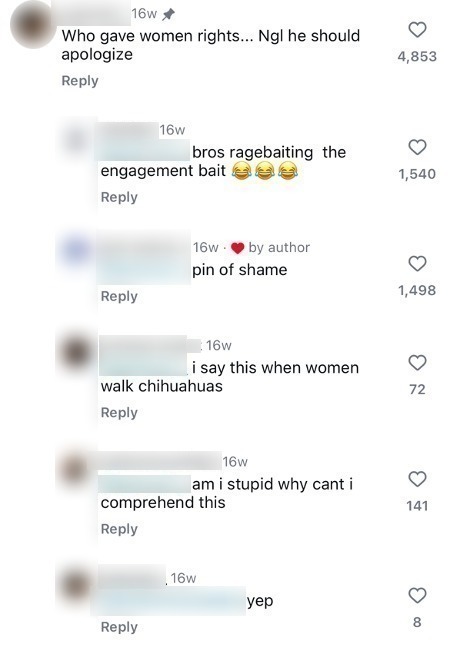}}
    \fbox{\includegraphics[width = 4cm]{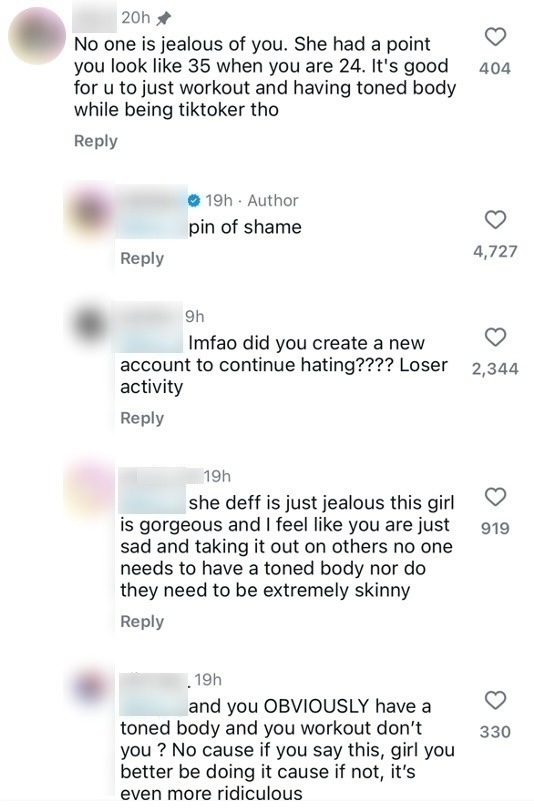}}
    \fbox{\includegraphics[width = 3.95cm]{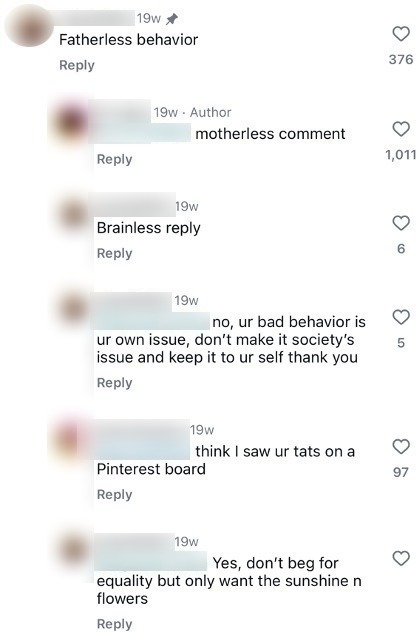}}
    \caption{Examples of Pin of Shame found on Instagram. Content creators pin the negative comments to fix them at the top of the comments section---either without posting any additional reply themselves (left), explicitly inviting their viewers to counteract the pinned comment (center), or mocking the pinned commenter (right).}
    \label{fig:example}
\end{figure}

Against this backdrop, we examine in this paper how, why, and to what effect content creators like Jayan deploy the `Pin of Shame'---a strategic pinning of inappropriate comments to invite shaming comments against them---to address content-based harm. 
This emergent regulation approach is notable in that platforms do not explicitly present comment pinning as a content moderation mechanism.
Across social media platforms, pinning is typically introduced as a visibility feature, allowing users to keep selected content at the top of profiles, chats, or comment sections.
%---creators are free to pin (i.e., force fix) any comment they choose to the top of their post's comments section so as to afford it greater public visibility. 
Platform descriptions often frame pinning as a way to emphasize preferred or positive content, rather than to regulate inappropriate behavior~\cite{youtube_pin,instagram_pin}. 
Yet creators on multiple platforms are increasingly deploying it to induce social sanctions against a range of norm violations, including identity-based attacks (as in the case of Jayan) and personal insults (see Figure~\ref{fig:example}).
In some cases, creators pin the inappropriate comment and leave no further reply (e.g., in Figure~\ref{fig:example}, left), while in other cases, they explicitly signal its inappropriateness (and hence, invite public shaming) by remarking ``Pin of Shame'' (Figure~\ref{fig:example}, middle), or clapping back with witty or sarcastic replies (Figure~\ref{fig:example}, right).

This recurring (yet understudied) use of the Pin of Shame suggests that platform-offered moderation tools may not be addressing some of the key creator goals for handling norm violations. Therefore, we take a creator-centered approach to understanding the deployments of the Pin of Shame. Doing so allows us to scrutinize creators' situated experiences of online harms~\cite{scheuerman_framework_2021,scheuerman_safe_2018}, how they perceive the tradeoffs between different coping mechanisms, and how they rely on their community for both emotional and social needs.
We also evaluate how the effects of this communicative form of social sanctioning speak to recent calls for developing alternatives to the current punitive moderation models—alternative approaches that emphasize acknowledgment, responsibility, emotional engagement, and, when possible, apology from rule violators~\cite{schoenebeck2021drawing,xiao2023addressing}.

%These varying examples of the Pin of Shame 
%draws from two key dimensions: the socio-technical affordances of the comment pin feature and the emotional dynamics of online shaming. 
%Prior research shows that moral outrage can trigger collective action in online networks~\cite{brady2017emotion,lewis2021we}.
%Content creators may leverage this dynamic by strategically highlighting inappropriate comments through pinning, aiming to provoke a public response.
%Collective participation lead by online shaming serves as a form of emotional punishment, evoking embarrassment and remorse~\cite{shea2024discursive}. 
%Recognizing these affordances and emotional effects, creators re-purpose the pinning feature to perform a symbolic form of moderation.
%
%Building on this conceptualization, we explore the contexts in which content creators adopt the Pin of Shame.
Although platform-defined community guidelines often shape users’ understanding of moral transgressions~\cite{kou2021flag}, the wide variety of norm violations represented in the Pin of Shame examples reflect how content creators may be perceiving and interpreting content inappropriateness within their own ethical frameworks. 
Further, by re-purposing the pinning feature---typically used to spotlight informative comments~\cite{hao2024shaping}---to instead highlight negative ones, creators are \textit{inverting the platform’s sanctioning logic of reducing visibility as punishment} ~\cite{gillespie2018custodians,schoenebeck2021drawing}. 
This practice signals creators' own vision of what comment moderation and sanctioning practices should look like. To more deeply understand what drives content creators to adopt the Pin of Shame as a coping mechanism, especially given the availabililty of alternative platform-offered moderation solutions (e.g., comment removal, reporting, blocking), we ask:

\textbf{RQ 1: What motivates content creators to use the Pin of Shame in response to comments perceived as inappropriate or norm-violating?}

%Online shaming is frequently portrayed in academic discourse as a negative practice, and in some cases, to escalate into offline consequences for the shamed individual~\cite{}.
%Second, drawing on the understanding that collective shaming actions contribute to a sense of community belonging~\cite{billingham2020enforcing}

By securing the visibility of inappropriate comments, the Pin of Shame draws users into a collective act of shaming the pinned commenters.
This form of collective social sanctioning is novel in that content creators intentionally leverage their audience's emotional involvement as a key procedural driver of their moderation practice.
Collective action based on moral outrages is often portrayed negatively, due to its potential to inflict mental and social harm to the shamed individuals~\cite{muir2021portrayal,basak2016look,basak2019online,ronson2015so}.
Our study seeks to complicate this framing by examining the perspectives and experiences of content creators who instigate online public shaming, especially attending to the benefits, uncertainties, and risks they perceive of this practice. 
We also explore whether and how deploying the Pin of Shame offsets the drawbacks of visibility-limiting moderation practices (e.g., comment removal), such as the undermining of free speech~\cite{myers2018censored} and opaque decision-making processes~\cite{jhaver2019did,suzor2019we}.
We ask:

\textbf{RQ 2: What are the consequences of using Pin of Shame in response to comments perceived as inappropriate or norm-violating?}

% The Pin of Shame, which visibility-based social sanctioning, is exactly opposite to the current practices of moderation, which aim to limit the visibility of harmful content to minimize its impact~\cite{gillespie2018custodians,schoenebeck2021drawing}.

%Additionally, the lack of transparency and accountability in such punitive actions has prompted calls for more reflective and restorative approaches to moderation~\cite{shahid2023decolonizing,kou2021punishment}.reporting~\cite{kou2021flag,crawford2016flag,wohn2019volunteer}, content creators choose to the Pin of Shame to achieve their specific motivations.
% Further, this communicative form of social sanctioning aligns with recent calls for more restorative approaches to moderation—emphasizing acknowledgment, responsibility, and, when possible, apology from rule violators~\cite{schoenebeck2021drawing,xiao2023addressing}.
%The Pin of Shame may offer a way to avoids current criticisms and minimize backlash from users, and nonetheless produce the intended behavioral and communicative outcomes.
% Thus, we aim to explore these aspects by comparing the limitations and benefits of the Pin of Shame with those of other available moderation tools.
% Accordingly, we pose the following questions:

% \textbf{RQ 3: What role does comment pinning play in the ecology of content moderation, and what values does it reveal about moderation?}
\vspace{5pt}
\noindent
To address these research questions and gain a nuanced understanding of the Pin of Shame phenomenon from the perspective of content creators, we conducted semi-structured interviews with 20 social media creators who had used comment pinning as a strategy to address inappropriate comments.
% 
%Findings+Discussion
Conducting a reflexive thematic analysis~\cite{braun2006using,byrne2022worked} of our interview data, we found that content creators adopt the Pin of Shame to draw public attention as a means to shame the offenders as well as shape community norms and educate viewers about those norms.
Some creators also seek to leverage the Pin of Shame to increase engagement with their posted content, operating on their own (potentially incorrect) folk theories about how algorithmic recommendations operate.
%By socially sanctioning inappropriate comments, 
Our findings show that using the Pin of Shame allows creators to make sense of the harm, validate their emotional responses, receive social support, hold norm violators accountable, and feel empowered.
Moreover, its use enables creators to avoid direct censorship and uphold free speech while reinforcing their positive public impression and strengthening their control over the unfolding discourse.
%These findings shed light on the multifaceted role of online shaming, not just as a punitive act but as a socially meaningful mechanism enabling norm enforcement, collective sanctioning, and ongoing norm negotiation. 
These results support prior scholarship that has explored the constructive potential of online shaming as a tool for norm enforcement~\cite{braithwaite1989crime, corry2021screenshot, direk2020politics, hou2017socioeconomic, kasabova2021shame, scheff2014ubiquity, trottier2018coming}.
However, our findings also reveal how the uncalibrated nature of public shaming can cause emotional strain for the instigators (in this case, content creators) themselves.

%By foregrounding emotionally accountable and restorative approaches to content governance, our study contributes to the fields of HCI and CSCW.
Prior HCI and CSCW research has examined the content suppression and moderation that content creators themselves experience on social media platforms~\cite{ma2023defaulting,ma2021howad,ma2023howdousers,harris2023black,register2024beyond,delmonaco2024you,kojah2025dialing}. We build on this work by contributing to the strand of research that evaluates how creators manage the user-generated content within their regulation purview, specifically the comments their posts receive online~\cite{jhaver2023decentralizing,jhaver2022designing}.
Our work adds to a growing body of content moderation scholarship that emphasizes the design of tools, techniques, and strategies aimed at empowering end users to combat content-based harms~\cite{blackwell2018online,im2020synthesized,jhaver2018online,mahar2018squadbox,vitak2017identifying,jhaver2023personalizing}.
Our investigation also responds to recent calls from HCI researchers to interrogate the role of emotion and discursive approaches in enacting content moderation~\cite{blackwell2018online, kou2021punishment, schoenebeck2021drawing, xiao2023addressing, french2017s, gelman2011concepts}.
Further, we extend prior conversations on incorporating accountability and transparency in moderation solutions~\cite{myers2018censored, zhang2023cleaning,shim2024incorporating,vaccaro2020end}.

% Meanwhile, deploying the Pin of Shame reflects users' value for greater transparency and norm adjustment in moderation, offering a response to the opacity and rigidity of current systems~\cite{myers2018censored, zhang2023cleaning}. 
% This finding contributes to the CSCW field by demonstrating how discursive-driven, collective participation may address governance challenges, advancing the discourse on fair moderation practices that facilitate norm adherence.
% Finally, our research adds value by suggesting design implications---another important aspect of CSCW research---such as preventing collective harassment toward norm violators, and developing emotion-sensitive algorithms that affect to the harmed users' behavior~\cite{french2017s, gelman2011concepts}.

%considering the nature of social media, which facilitates peer surveillance through public participation~~\cite{muir2021portrayal}.

%what is punishment, and what is counter-narrative~\cite{shea2024discursive}.
%example of online shaming : https://www.nytimes.com/2015/02/15/magazine/how-one-stupid-tweet-ruined-justine-saccos-life.html

\section{Related Work}

This study draws from two related areas of research: content moderation practices and online public shaming action in response to perceived moral transgressions. 
Together, these research areas inform our analysis and provide the context for how content creators creatively leverage the comment pinning feature to manage and respond to harmful comments.

%\subsection{Creator-led Content Moderation}
\subsection{Content Moderation and Creators' Moderation Practices}
%Moderation definition

\subsubsection{Different Manifestations of Content Moderation.}
Content moderation is broadly defined as a set of sociotechnical practices that seek to identify and address inappropriate user-generated content on social media sites~\cite{gillespie2018custodians,roberts2019behind}.
These practices include, but are not limited to, creating codes of conduct, reviewing posts for norm violations, implementing automated detection, sanctioning undesirable behaviors, and changing the visibility or reach of inappropriate content.
On most mainstream platforms, content moderation occurs at multiple levels of governance~\cite{jhaver2023decentralizing}. That is, platform admins hold the power to sanction any posts, users, channels, or communities site-wide, whereas end-users also have the ability to mold the type of content they see through configuring what has been termed as \textit{personalized moderation} tools, such as word filters and toxicity sliders~\cite{jhaver2023personalizing}.
Further, at the middle level of governance between platforms and end-users~\cite{jhaver2023decentralizing}, entities such as community managers (e.g., moderators of subreddits and Facebook Groups) and content creators (e.g., YouTube channel managers) can regulate content posted under their purview (e.g., posts within their community, replies to their YouTube video), but are themselves also accountable to platform-enacted regulation.

% Content moderation actions aim to enforce norms of appropriate conduct. 
% These norms are collectively shaped by platforms, algorithms, creators, and end-users, and they are influenced by--but are distinct from--offline societal standards.
% Given the decentralization of governance authority on platforms (as discussed above), these norms operate at different scales~\cite{chandrasekharan2018internet}: certain posts (e.g., child pornography) are unacceptable site-wide, whereas other actions (e.g., sarcastic comments, personal opinions) are acceptable on some spaces but regulated elsewhere.  
% % Further, norms of conduct vary widely between units at middle level of governance. 
% For instance, content creators have diverse expectations of the type of replies that are acceptable to post on their channel. These expectations reflect among other factors a mixture of different cultural backgrounds and site usage histories. Creators' moderation attempts often seek to align user comments on their channel in accordance with their expectations. 

The enactment of moderation--either by platform admins, community managers, or channel owners--represents an exercise in authority to determine which speech and whose speech deserves to be sanctioned~\citep{gillespie2018custodians,suzor2019lawless}. Prior research has examined such power dynamics between different levels of governance, e.g., between platforms and communities~\citep{jhaver2025bans,chandrasekharan2017you}, platforms and end-users~\citep{myers2018censored}, community managers and end-users~\cite{matias2019civic,jhaver2019did} and creators and end-users~\citep{jhaver2022designing}.
We attend to such power relations in the context of comment pinning practices in this study.

% It is an exercise of power: it encompasses deterrence (preventing harm) and punishment (responding to harm) actions, regulating what actions are permitted, and how norms are internalized within online communities~\cite{grimmelmann2015virtues}. 
% Through such mechanisms, users learn what behaviors are acceptable and how they are expected to conduct themselves online~\cite{grimmelmann2015virtues}.

\subsubsection{Creator-led Moderation}
The current paper is concerned with creator-led moderation practices. 
Content creators on digital platforms routinely encounter harmful comments from other users, either as responses to their posted content or elsewhere on their news feeds~\cite{harris2023black,jhaver2022designing}.
Such comments include instances of online harassment~\cite{schoenebeck2021drawing}, hate speech, graphic content, and misinformation~\citep{shea2024discursive}.
In response, creators employ a range of coping mechanisms~\cite{thomas2022s}, including punitive measures (e.g., flagging, blocking, deleting~\cite{kou2021flag,crawford2016flag,john2024classification,jhaver2018online}), visibility-reduction tools (e.g., toxicity sliders and keyword filters~\cite{jhaver2022designing,jhaver2023personalizing}), and discursive interventions like counter-speech~\cite{blaya2019cyberhate,blackwell2018online,shea2024discursive}.

%Mod measures have diff motivations and effects
These mechanisms vary in their accessibility, granularity of control, scope of impact, effectuation delays, and certainty of outcomes~\cite{jhaver2023personalizing,zhang2023cleaning}. 
For instance, flagging a comment aims to remove it platform-wide for all users, but its impact is delayed and contingent on platforms' flag review decisions~\cite{crawford2016flag}.
In contrast, muting or keyword-based filtering hides content only for the configuring creator, and its effects are certain and immediate~\cite{jhaver2023personalizing,jhaver2022designing}.
On the other hand, \textit{disliking} a comment does not remove or hide it, but publicly indicates disapproval---the impact is immediately visible and the dislikers usually remain unrevealed~\cite{lee2022negative}. 
Given these differences, content creators often tailor their choice of response to inappropriate behaviors to achieve their specific goals~\cite{chen2013and, thomas2022s, masullo2018nastywomen}. 
Within this ecosystem of creator-led moderation mechanisms, the `Pin of Shame' has emerged as a novel form of creator-led moderation to address online harm, as we next describe. 

\subsubsection{The ``Pin of Shame'' as a Creator-led Moderation Mechanism}

While the comment pinning feature was originally designed to highlight valued comments, creators have creatively repurposed it for various goals, such as monetizing content or disclosing sponsorships~\cite{hao2024shaping}. 
Most notably for the purpose of this study, some creators now use it as a mechanism of social sanction, leveraging the visibility control it provides to publicly call out norm violations. 
Our study focuses on this specific use of comment pinning.
Given that the use of Pin of Shame \textit{changes the visibility} of inappropriate content, it falls under the scope of prior conceptions of content moderation mechanisms~\cite{gillespie2018custodians,roberts2019behind,grimmelmann2015virtues}, and we conceptually treat it as such.
Indeed, \citet{lambert2024positive} have similarly framed another appropriation of comment pinning as moderation: they note that Reddit moderators pin (``sticky'' in Reddit vernacular) desirable posts and comments to boost visibility, theorizing this positive reinforcement as a significant moderation practice.
% and emphasized how moderators deploy the sticky feature on Reddit to control visibility---this further supports our conception of comment pinning for shaming as a moderation mechanism.
% , examining how they deploy the Pin of Shame as a mechanism of punishment and norm education.
% As our findings will also concur, in our participants' emic perspectives, deploying the Pin of Shame serves as a form of user-driven sanction~\cite{jhaver2023personalizing} that retains the inappropriate comment while directing its strategic visibility into a mechanism of punishment and norm education.
The Pin of Shame also aligns with retributive justice, wherein people who commit moral transgressions receive a proportional sanction for their misdeeds~\citep{blackwell2018online}. We will use this case study to theorize shame as a justice practice in platform governance.

It is difficult to ascertain how widespread the Pin of Shame use is across platforms. This is because platforms do not treat this practice as a discrete intervention or provide any usage statistics about it.
Besides, over the past few years, comprehensively collecting social media data is becoming increasingly prohibitive ~\citep{davidson2023platform,jhaver2023GettingData}, so counting discrete instances of this practice is challenging.
However, the Pin of Shame presents an instrumentally unique addition to the toolkit of moderation interventions available to creators. Thus, it presents a valuable case study to examine how creators perceive, compare, select, and deploy different approaches to content moderation in response to specific infractions and to achieve distinct punitive and educational goals~\cite{jiang2023trade,myers2018censored}.

%Norm-enforcement & CSCW contribution

\subsubsection{Creators’ Impression and Visibility Management}
Content creators engage in building and maintaining relationships with their viewers~\cite{hamilton2014streaming,hilvert2018social,kim2019will,wohn2020audience}. 
These engagements are driven not just by creators' communal aspirations but also their self-promotion goals and platforms' commercial logic~\cite{boxman2019practice}.
Many creators adopt ``micro-celebrity'' practices, treating their audience as fans, framing content creation as a performance~\cite{li2019live,pellicone2017game}, and leveraging strategic intimacy to connect with viewers through direct engagement~\cite{berryman2017guess,marwick2015you,raun2018capitalizing}. 
% The technical features of social media platforms—such as options to view, comment on, and share—further facilitate the formation of dynamic, networked communities~\cite{cunningham2017being,lange2007commenting}, where creators and audiences actively engage with one another~\cite{burgess2018youtube}.
Given the importance of building and maintaining relationships with viewers in creators' work~\cite{hamilton2014streaming,hilvert2018social,kim2019will,wohn2020audience}, the hyper-visibility of moderation offered by the Pin of Shame may open up opportunities for audience engagement and relationship-building.
Unlike visibility-reduction tools—whose hidden and unilateral actions can strain audience trust—visibility-enhancing sanctions may invite public engagement by encouraging further interaction around the pinned comment.

We examine in this research how creators' audience engagement and impression management goals align with and influence their comment moderation efforts through their use of comment pinning alongside other conversation 
affordances, such as the ability to reply and remove comments.
Many HCI scholars have attended to how creators seek to understand the way that recommendation algorithms work on these platforms so as to enhance their visibility and subsequently profit~\cite{cotter2019playing,ma2021howad,karizat2021algorithmic,devito2022transfeminine}. Building upon this, we inquire how creators' mental models of recommendation algorithms shape their use of comment pinning for norm enforcement.

\subsubsection{Labor Demands of Creator-led Moderation}
Previous studies have raised significant concerns about the physical and emotional toll that enacting content moderation takes on various groups, including regular social media users~\cite{jhaver2023personalizing,myers2018censored,haimson2021disproportionate}, commercial moderators~\cite{roberts2019behind, steiger2021psy}, volunteer moderators~\cite{li2022all, jhaver2019automated, wohn2019volunteer, dosono2019moderation, ruckenstein2020re}, and---most relevant to our study---content creators~\cite{jhaver2022designing,ma2022m,meisner2023weaponization,heung2024vulnerable}. Researchers have also highlighted the tensions surrounding how platforms profit from users’ unpaid moderation efforts~\cite{wohn2019volunteer, li2022measuring}. We add to this literature by examining how the use of Pin of Shame is a reaction to these cognitive and emotional labor demands, and how creators leverage their community's support to ameliorate these demands.

\subsubsection{Transparency and Control in Moderation}

Currently, most social media platforms provide limited insight into the rules and decision-making processes that underlie platform-enacted moderation practices~\cite{suzor2019we, jhaver2023decentralizing}. 
% Even user-controlled moderation tools that rely on platforms' adjudication (e.g., flagging~\cite{crawford2016flag}) or curation algorithms (e.g., toxicity sliders~\cite{jhaver2023personalizing}) often present a lack of transparency in their procedures, interface elements, and outcomes. 
% 
This lack of transparency negatively impacts user experiences.
Prior studies show that visibility-reducing actions like comment deletion are frequently criticized for undermining free expression and provoke backlash from audiences~\cite{jhaver2022designing,ma2023defaulting}.
Because such actions are often unilateral and opaque to observers, they can erode their perceived sense of transparency and procedural fairness.
On the other hand, when moderation systems enact transparency in their operation, e.g., when they explain the rationale behind content regulation decisions, sanctioned users respond more positively, leading to improved post quality~\cite{jhaver2019does} and greater trust in the platform~\cite{jhaver2019did, brunk2019effect}. 
More importantly, such explanations also positively shape other bystanders' behaviors~\cite{jhaver2024bystanders}. 
We examine in this article how the transparency afforded by pinning the shamed comment (as opposed to silently removing it) influences observers' reactions to comment sanction in the view of creators.

Further, in designing moderation systems, it is crucial to account for end-users’ preferences for control~\cite{ozanne2022shall, juneja2020through, suzor2019we}. 
Researchers have explored how moderation tools that clarify the definition of interface elements and incorporate the cultural context of posts can help enhance users’ perceived control over their social media feeds~\cite{jhaver2022designing, jhaver2023personalizing}.
We attend to how content creators' need for control influences the specific ways in which they engage in comment pinning for moderation, e.g., pinning and unpinning different comments and following up pinning with comment removal.
In doing so, we add to ongoing HCI conversations around empowering end-users with shaping custom solutions to address online harm~\cite{vaccaro2020end,shim2024incorporating,jhaver2019did}.

\subsubsection{Creators' Authority to Shape Norms}
Previous research on the use of personal moderation tools like blocking, word filters, and toxicity sliders shows that user-controlled interventions enable the assertion of user-defined norms of appropriateness~\cite{jhaver2023personalizing,jhaver2018online,jhaver2022designing,luther2026moderation}---this contrasts with tools like flagging that are constrained by platforms' narrow vocabulary of inappropriate conduct~\cite{kou2021flag,crawford2016flag,zhang2023cleaning}.
We examine how using the Pin of Shame allows creators to precipitate a process of collective norm negotiation, co-construction, definition, and enforcement among their audiences.
We analyze how the Pin of Shame operates as a discursive tool that invites public reflection and user interaction, aligning with emerging CSCW research that explores restorative justice frameworks and community-based responses to online harm~\cite{xiao2023addressing,schoenebeck2021drawing,blackwell2018online,asad2019prefigurative}. 
% We analyze how this practice foregrounds the agency of everyday users in shaping content moderation discourse and practice. 

Prior work shows that creators often wield substantial social influence over their audiences, which derives from the enhanced attention they receive (as compared to other commenters in their channel) and the relationships they develop with their followers~\cite{barari2023unveiling,ekinci2025dark,harff2022responses,liu2024persuasive}. 
Indeed, creators’ personal impulses, commercial incentives, or malevolent intentions can lead them to promote or enact inappropriate norms on their channel~\cite{barari2023unveiling,ekinci2025dark,levin2025influencer}.
Thus, creators' behaviors can have harmful effects, e.g., they can boost the spread of misinformation or expose their followers to beauty ideals that induce appearance-related anxiety~\cite{harff2022responses,levin2025influencer}.
% When their oversized authority is combined with  amplify potential harms when creators act on non-benevolent motivations. 
%At the same time, creators’ desire for greater control over their comment spaces shapes how they enact moderation practices, particularly in contexts where platform-enacted moderation is opaque or insufficient.
Given the strong authority that creators have within their channel, we explore how the Pin of Shame can be used inappropriately and precipitate potential harms via creators' influence.

\subsection{Online Shaming as a Norm Enforcement Mechanism}
\subsubsection{How Public Shaming Manifests Online}
\textit{Shame} is a self-conscious social emotion~\cite{jochan2021shaming,scheff2014ubiquity,tangney1995shame} often associated with embarrassment~\cite{shea2024discursive} and the experience of feeling small, worthless, or powerless~\cite{tangney1995shame}. 
On networked social media platforms, many norm-violating acts are publicly visible, increasing the likelihood that they spontaneously elicit a \textit{shaming} response~\cite{de2015use,jochan2021shaming,tangney1995shame}. 
In other cases, online shaming is deliberately initiated by uploading videos, photos, or screenshots of norm-violating behaviors to attract public criticism~\citep{corry2021screenshot,hou2017socioeconomic}, publicly listing users who have breached community norms~\cite{schoenebeck2021drawing}, or calling for specific violators to be held accountable.
Many affordances of popular social media platforms, such as content persistence, searchability, and shareability facilitate online public shaming~\cite{kim2022does,jhaver2018online}.

Such online shaming manifests as witnesses collectively condemning the norm violator's actions and calling for accountability and behavioral change~\cite{de2015use,scheff2014ubiquity,muir2021portrayal} in response to their perceived moral transgression~\cite{basak2019online,nussbaum2019shaming}.
These responses are often driven by in-group dynamics, as individuals may interpret norm violations---particularly by out-group members---as symbolic threats to their collective identity~\cite{stets2000identity}.
Such shaming actions help define the boundaries of acceptable conduct~\cite{hou2017socioeconomic}, educate the shamed users about the consequences of norm violations, and reinforce the internalization of community norms among observing individuals~\cite{trottier2018coming,corry2021screenshot}.
Further, in repressive or authoritarian contexts, online shaming may serve as a tool for challenging existing power structures~\cite{direk2020politics}.

\subsubsection{The Pin of Shame as a Unique Instantiation of Online Public Shaming}

The Pin of Shame enables a unique creator-directed and follower-mediated form of public shaming where the original infraction, its visibility enhancement (through comment pinning), and the subsequent shaming (and counteractions to shaming) all occur within the same deliberative space.
From the point of view of creators, shaming induced by the Pin of Shame (in contrast to a top-down platform-enacted sanction) may hold significant potential as a form of grassroots content moderation~\cite{kasabova2021shame}, where their community of audiences plays an active role in upholding social norms.
% Unlike other general forms of online shaming, this practice allows for ongoing interactions among the creator, target(s), and audiences.
Creators have significant authority in shaping the trajectory of Pin of Shame. For instance, creators can follow up pinning with counterspeech or comment removal.
Further, the active presence of targets increases the likelihood that they are aware of the ongoing shaming. It also allows the targets to take further actions, such as defending themselves, apologizing, or deleting their pinned comment.
We examine creators' perspectives on how these unique affordances distinguish Pin of Shame from other instances of online public shaming.

\subsubsection{Public Shaming as a Social Sanction versus Networked Harassment.}
%Negative aspect
% Shaming is usually characterized by a negative rhetoric.
% In some instances, online public shaming , inflicts emotional harm on the targeted individual, and devolves into harassment~\cite{jhaver2017view}, which perpetuates cycles of punitive outrage~\cite{basak2016look,marwick2021morally,lewis2021we}.
Prior research has documented how online public shaming can cross moral boundaries~\cite{billingham2020enforcing}, escalate into networked harassment~\citep{blackwell2017classification,marwick2021morally}, and cause emotional harm against the targets while inducing them to self-censor~\citep {marwick2021morally,blackwell2018online,ronson2015so}.
For instance, \citet{marwick2021morally} highlights that morally motivated harassment is frequently driven by a desire to signal network membership and reinforce the network's values.
% Such shaming actions are often carried out beyond the context in which the initial shaming occurs~\cite{ronson2015so}, spreading across platforms and even into offline settings, where targets are deprived of meaningful opportunities to respond or defend themselves~\cite{lewis2021we, marwick2021morally, ronson2015so}.
Over time, shaming can escalate into identity-based accusations~\citep{marwick2021morally}, producing enduring stigma for those who are shamed~\citep{ronson2015so}.
% 
% This body of work highlights how online communities adopt public call-outs to enforce social norms, particularly where formal rules are ambiguous~\citep{fiesler2019creativity}.
%this raises the likelihood that shaming actions that seem ethically justified to a large group cause psychological damage such as depression and anxiety in the accused.
Building on this literature, our work considers how deploying the Pin of Shame speaks to the questions of boundary between harassment and justice-oriented shaming.
% shaming may function as a form of moderation in creator-led comment management, rather than devolving into harassment, through the case of the Pin of Shame.
%Such an outcome may not be desirable even for the creators.

Specifically, we examine shaming as it unfolds within a localized and bounded setting---the creator’s comment thread---where norm violations and responses remain visible, enabling situational awareness and defense by the targeted commenter.
Moreover, the affordances of comment pinning, such as the ability to unpin comments (from the creator’s side) or delete comments (by both creators and commenters), introduce reversible interventions that can curb shaming, whenever desired.
These distinctions suggest that the Pin of Shame participants assert a greater control over whether shaming escalates into harassment or remains oriented toward sanctioning and norm negotiation.
Accordingly, we examine how the choices made within the lifecycle of Pin of Shame deployments limit the scope of visibility and shape the dynamics of public sanctioning.

Further, prior research on norm enforcement highlights that shaming can take on an educational orientation~\citep{fiesler2019creativity}.
When shaming is used to label norm violations and articulate community standards~\citep{blackwell2017classification}, it can facilitate processes of reintegration rather than exclusion~\citep{braithwaite1989crime}.
Building on this work, we examine how norms are articulated, negotiated, and communicated through Pin of Shame interactions, and how these processes shape participants’ understandings of acceptable behavior.

\section{METHODS}

\subsection{Participant Recruitment}

%Our 
Rutgers University's
%\footnote{We will reveal the university’s name after the peer review process completes.} 
Institutional Review Board (IRB) reviewed the study and categorized its status as exempt on February 3, 2025.
We employed purposive sampling~\cite{campbell2020purposive} to select our interviewees, targeting only those content creators who had responded to negative comments by pinning them.
The comment pinning feature is not available on many social media platforms like Facebook and TikTok. 
Among the platforms that offer it, we recruited participants primarily from Instagram and YouTube, as they represent mainstream social media services. 
Our conversations with participants revealed that the ``Pin of Shame'' phenomenon occurs with notable frequency on these platforms, which further confirms their relevance as sites of participant selection for this study.

% \subsubsection{Subject Selection Criteria}
% \subsubsection{Recruitment Procedure}
Our initial interest in this phenomenon arose when the first author encountered several instances of creators pinning negative comments during her everyday use of Instagram and YouTube.
In these initial examples, we found that the pinned comments frequently targeted creators’ identities or viewpoints, particularly regarding gender, political attitudes, and religion.
Prior research also suggests that posts associated with topics like gender, disability, religion, and elections often invite disproportionate levels of harassment, hate speech, and misinformation~\cite{ADL2024,vogels2021state,marwick2018drinking}.
%We could not find any curated list of posts displaying the Pin of Shame.
Therefore, to retrieve a broader sample of the Pin of Shame posts we had coincidentally encountered, we searched Instagram and YouTube using targeted keywords such as ``LGBTQ,'' ``MAGA,'' and ``Christian'' which we expected to surface posts more likely to contain pinned negative comments 
We conducted these searches using study accounts created for Instagram and YouTube specifically for this study.
% To find potential interview subjects, 
% we created study accounts on Instagram and YouTube, and conducted keyword searches on both sites to retrieve posts with a higher likelihood of featuring pinned negative comments.

We manually reviewed the posts returned by these initial searches to determine whether they contained instances of the comment pinning feature being used for shaming. 
Filtering for such posts, we then examined the hashtags and terms commonly used in them to find additional relevant keywords.
For instance, we observed that some creators used the keyword ``PinofShame'' to relate their experiences with comment pinning as a means of shaming, and we added that keyword in our corpus search.
We also reviewed the posting history of some creators who prominently submitted Pin of Shame posts, and noted the topics and hashtags associated with those posts.
% When we identified additional instances of the Pin of Shame on the same creators' accounts, we also noted the topics and hashtags associated with those posts. 
We excluded keywords that were overly broad, unrelated to the focal practice, or consistently returned irrelevant posts.
Thus, this keyword building process was iterative and incremental, and while our initial keywords likely somewhat influenced our selection, our multi-pronged approach allowed us to extend our initial list by adding a range of keywords such as ``Feminist,'' ``Queer,'' ``Plussizelife,'' and ``Abortion.''
In total, we used 20 keywords for the search (see Table \ref{keywords}).

\begin{table}[h]
    \centering
    \small
    \renewcommand{\arraystretch}{1.2} % row spacing
    \setlength{\tabcolsep}{10pt} % column padding
    \begin{tabularx}{0.85\textwidth}{XXX}
        \noalign{\hrule height 1pt}
        \multicolumn{3}{c}{\textbf{Keywords Used to Find Pin-of-Shame Posts}} \\
        \noalign{\hrule height 1pt}
        Abortion & Christian & Disability \\
        Disorder & Feminism & Feminist \\
        Immigration & LGBTQ & MAGA \\
        Non-binary & PinofShame & Plussizelife \\
        Politics & Prochoice & Queer \\
        Transgender & Viralvideo & Womenempowerment \\
        WomeninMaleFields & Womeninamerica & \\
        \noalign{\hrule height 1pt}
    \end{tabularx}
    \caption{Keywords used for finding example posts engaging with Pin of Shame practices in Instagram and YouTube}
    \label{keywords}
\end{table}

Additionally, we liked and bookmarked posts displaying the Pin of Shame so that the platforms' feed recommendations could further assist our data collection. Following this strategy, we observed additional instances of Pin of Shame in our study accounts' news feeds.
% For each post, we documented the account, content, captions, pinned comments, and subsequent discussion.

Based on our review of the posts we collected, we identified 239 individuals as potential interview candidates.
For each of these individuals, we documented their account details. Additionally, the post where they deployed the Pin of Shame, its description, pinned comment, and replies to that comment were collected.
Next, we contacted candidates identified on Instagram using the site's direct message feature. 
On YouTube, we contacted candidates who had publicly listed their email-address on their channel profile by emailing them. 
During the interview recruitment phase, we also employed snowball sampling, with prior participants recommending other potential candidates.

\subsection{Interview Procedure}
In total, we conducted semi-structured interviews with 20 individuals. 
All interviews were conducted between February 4th to April 10th, 2025 via Zoom, and interviewees were allowed to join either via their computers or their phones.

The interviews consisted of four parts: background information on their
channel or profile activities; general strategies for addressing negative comments; motivations and outcomes associated with their use of the Pin of Shame---this is where we focused the bulk of our discussion; and perspectives on comment pinning as a content moderation strategy.
During each interview, our lines of inquiry and follow-up questions somewhat varied, depending on each interviewee’s experience, allowing us to probe further into participants' specific expectations, values, outcomes, and comparisons with other moderation affordances.
For example, if the interviewee had flagged inappropriate comments, the interviewer also asked additional questions, such as how that experience influenced their use of the Pin of Shame.

Before each interview began, we reviewed the data related to the Pin of Shame posts made by that interviewee, including the pinned negative comment and replies to that comment, to orient ourselves. 
During the interview, we referenced this data using screen sharing to guide our conversation with the participant. 
Some participants wanted to share their experiences regarding additional posts where they had deployed the Pin of Shame; we encouraged participants to discuss such cases with us.
The interviews were transcribed using Zoom's transcription feature, subsequently cleaned, and then transferred to and analyzed using NVivo (v.15), a qualitative analysis software.

We concurrently conducted our preliminary data analysis with participant recruitment. 
After the 15th interview, we began to observe theoretical saturation
~\citep{faulkner2017theoretical}, i.e., no new codes or categories were emerging, and all identified concepts in our analysis were well defined and supported by sufficient data.
To ensure the robustness of our findings, we recruited five additional participants and concluded data collection with a total of 20 interviewees.
% The interview duration ranged from 40 to 60 minutes, with the average duration being 47 minutes.
After each interview was completed, we compensated the participant with a \$20 Amazon e-gift card.

Table \ref{tab:demographic} (in Appendix \ref{sec:appendix-participants}) presents the demographic information of our participants.
Most of our participants were based in the United States.
Approximately 90\% of our participants were female, which is in line with prior findings that women are more likely than men to experience online harassment on social media and therefore use content moderation mechanisms~\cite{pew2021harassment,marwick2018drinking,jhaver2018online}.
Participants included a range of mid-size to large content creators (follower counts ranging from 1k to over 100k), producing content on diverse topics ranging from fitness and travel to social justice issues.
We conducted interviews only with users who could communicate in English or in other languages in which we are fluent.

\subsection{Data Analysis}
We addressed our research questions within a paradigmatic framework of interpretivism and constructivism~\cite{schwandt1994constructivist,byrne2022worked,clarke2014thematic}, i.e., we sought to aggregate our participants' individually constructed accounts of their motivations, attitudes, and experiences regarding their use of the Pin of Shame as faithfully as possible,\footnote{Given this emphasis, our analysis potentially under-reports the exploitative uses and harms caused by the Pin of Shame due to participants' desirability bias. This is a limitation of our study that we further comment on in Sec. \ref{sec:limitations}.} while still accommodating the reflexive influence of our own interpretations.
Given these goals, we analyzed our data using Braun and Clarke's approach to reflexive thematic analysis~\cite{braun2006using}.

We adopted an experiential orientation to data interpretation and an inductive approach to analysis to emphasize meaning and meaningfulness as ascribed by our participants~\cite{clarke2014thematic}.
We began by reading transcripts multiple times to familiarize ourselves with the interview data. 
The first author created open codes for each interview and wrote memos to document emerging insights.
% Throughout the transcript analysis, new codes and themes continued to emerge up to the 15th transcript.
% The primary researcher determined that theoretical saturation had been reached, as the 16th transcript yielded no new codes and could be fully coded using the existing code set.
Next, through an iterative coding process, we refined and consolidated the names and descriptions of codes, identifying  98 initial codes.
Throughout the coding process, the two authors discussed the selection of codes, the relations between codes and data, and how these codes relate to our research questions. 
We utilized both semantic and latent coding, but prioritized semantic codes.
Our coding was grounded in a constructionist epistemology. Specifically, we acknowledged the relevance of recurrence, but used our participants' expressed degrees of importance to the issues discussed and meaningfulness as interpreted by us as the central criteria for creating and assigning codes.

Following this, we visualized and developed the thematic connections between our codes.
To facilitate this, we physically wrote our codes on sticky notes and placed them on a wall --- this process offered us greater freedom to engage in constant comparisons, rearrangements, and reflections~\cite{maher2018ensuring}.
Using the 98 initial codes, we merged similar concepts and organized them into a hierarchical structure, represented by a set of new parent codes and corresponding child codes.
% This visualization and reorganization process enabled us to facilitate the emergence of a cohesive narrative from the data.
Through this process, we developed 12 parent codes and 76 child codes.
For instance, the parent code `Fostering norm education' included the child codes `educating bystanders' and `preventing similar actions in the future.'
Finally, after further reviewing and refining these codes and examining the relations between them, we developed thick descriptions of parent codes, and names them as our key themes that we present as our findings.
\section{Findings}

In this section, we first describe content creators' motivations, considerations, and concerns when deploying the Pin of Shame. Next, we discuss the social effects of deploying the Pin of Shame and how creators perceive these effects. 
%\begin{tcolorbox}
%CONTENT WARNING: This section contains offensive language, including misogynistic slurs, that readers may find disturbing.
%\end{tcolorbox}

\subsection{Motivations for Deploying the Pin of shame}
%Affordances and Motivation
\subsubsection{Inviting Public Attention to Invoke Shaming and Foster Norm Education}
\label{sec:Public attention}

\textit{Pinning} a comment places it at the top of the comments section. Understanding this function of the pinning affordance, most of our participants pinned negative comments to make them highly visible and more likely to attract other users' attention. 
%Rather than simply highlighting these comments for visibility, creators used the Pin of Shame to make the violation publicly legible, surface the community’s disapproval, and hold the commenter accountable in front of their audience.
%These accounts indicate that creators themselves understand pinning as a tool that can function as social sanctioning, even when they also use it for other purposes—such as humor, self-presentation, or engagement optimization—discussed in the following section.
This was often done to \textbf{raise their audience's awareness} of the harmful comments they face. For instance, Participant P5, a transgender woman who often receives transphobic comments on her posts, explained that she uses comment pinning to visibilize the everyday discrimination against transgender people that has become so normalized that it often goes unnoticed:

% Most of the time, content creators desire to \textbf{make negative comments highly visible to attract attention from other users}.
% Since  making it highly visible, creators believe it draws more attention, raising awareness of such negative comments and warning others about the harmful impact.

\begin{quote}
    \textit{``I pinned the comment because I just wanted you to see what people like me are going through in this country right now, like what people feel comfortable saying to us.'' -- P5}
\end{quote}

Beyond raising awareness, many participants also expected other users who see the Pin of Shame to engage in criticism or rebuttal against the pinned commenter. 
% This strategy is often successful---the Pin of shame frequently prompts comment viewers to both post \textbf{counterarguments against the negative commenter} and \textbf{defend the original poster (OP)}, transforming the comments section into a discursive space to confront, challenge, and correct inappropriate behavior.
% 
%\begin{quote}
%    ``I expected people being in my side. Yeah, more like, `she's right', or something like that. And actually, it turned out like that also, like people more. And my side, not by his side.'' - P1
%\end{quote}
Some saw this public criticism as a means to \textbf{humiliate and embarrass the negative commenter}. In this way, this strategy serves as a form of punishment.
For example, Participant P18 described her Pin of Shame action as such: 

\begin{quote}
    \textit{``I think punishment is a good word [to describe the Pin of Shame], just like it's public shaming. If you feel that strongly to say really mean stuff on a stranger's post, I think you should publicly be seen for that comment. You don't get to hide within the thousands of comments.'' -- P18}
\end{quote}

Thus, our participants' emic views of Pin of Shame deployments align with scholarly conceptions of what content moderation entails~\cite{gillespie2018custodians,grimmelmann2015virtues,kiesler2012regulating}. We, therefore, theorize this practice as a moderation invention.
Similar to conventional moderation, our participants shared that pinning allowed them to sanction comments that fell into categories such as LGBTQ denial, misogyny, anti-immigrant rhetoric, calls for violence against specific communities, or ethnicity-based hate speech.
%They believed those comments should not be allowed in the comment sections of their posts.

In addition to sanctioning, participants view the Pin of Shame as a tool to \textbf{educate the pinned commenters}, encouraging them to reflect on their actions, and prompting changes in their attitudes or future behaviors.
Participant P15 illustrated her expectation of using comment pinning against troll accounts: 
    
%\begin{quote}
%    ``So, if anything, pinning the comments will help them, maybe because then other people will argue with them and tell them they're a different perspective, and then maybe they can learn something from it.'' -P17
%\end{quote}

\begin{quote}
    \textit{``So if somebody has a troll account, and they make a nasty comment, and it gets pinned. And they have all these people talking back to them. The next time they decide to leave a comment on another creator's video, they're gonna think twice. `Is that person gonna pin my comment? And am I gonna get shamed again?' '' -- P15}
\end{quote}

In addition to reforming the offending commenter, participants expected that the Pin of Shame would also \textbf{educate bystanders}, i.e., people who witness the ensuing discussion, about the importance of adhering to social norms by publicly demonstrating the consequences of norm-violating behavior.
Participants hoped that observing these effects would discourage bystanders from posting similar comments on their channel. For instance, P3 mentioned:

\begin{quote}
\textit{     ``When I pin a comment, I feel like it can influence many people. So it’s not just about me managing the situation and moving on. Others can see it and think, `Oh, this is real.’ Some people who came to leave a negative comment might change their minds. I believe it can create secondary and even tertiary effects.'' -- P3}
\end{quote}

% Several participants also noted that the comment pinning feature is technically \textbf{easy to adopt, intuitive to use, and brings immediate results}, making it an appealing option for them to address negative comments with minimal effort.
% For example, P10 said:

% \begin{quote}
%     \textit{``I think that pinning would probably do more than reporting something, because it has more of an immediate effect on the commenter and stuff, where it's like, people are seeing you [the pinned commenter], and they're seeing that you're saying these things.'' -- P10}
% \end{quote}

\subsubsection{Shaping What Constitutes Norm Violations}
\label{sec:what constitutes norm}
Using the Pin of Shame gives creators the \textbf{freedom to decide what should be deemed a norm violation} and which comments deserve to be shamed and criticized on their channel.
Participants recognized that this flexibility allowed them to sanction comments ranging from those that seem ``ridiculous,'' ``judgmental,'' or ``condescending,'' to more serious infractions, such as discriminatory statements, misinformation, and hate speech.
% It does not require considering platforms' vocabulary of undesirable content. 
In contrast, flagging constrains regulation by requiring creators to select an inappropriate content category (e.g., `Violent Content', `Intellectual Property') under which the flagged post violates the platform's code of conduct~\cite{crawford2016flag,are2023flagging,zhang2023cleaning}.
Participants also appreciated the control over outcomes that the Pin of Shame offers, especially when compared to flagging, where outcomes (e.g., content removal) depend on whether the platform regulators consider the flagged content a violation.

While participants appreciated not being limited by platform guidelines when deploying the Pin of Shame, they also recognized that its \textbf{success still depends on the moral outrage} of the audience and their subsequent pushback against the pinned comment. 
In other words, while the authority to initiate the Pin of Shame and define the initial moral framing is concentrated in creators’ hands, the subsequent trajectory and consequences of the sanction are mediated through audience reactions.
As such, the Pin of Shame does not simply impose a fixed norm, but opens a space in which norms may be contested, reinforced, or redirected through public engagement.

Thus, some participants found it challenging to determine which negative comments to pin such that they would attract a broad public scrutiny and subsequent criticism. 
% While this control over outcomes allowed participants the flexibility to deploy the Pin of Shame against any comment, they noted the challenge of determining which negative comments deserve this treatment, since what is considered a negative comment is often subject to individual interpretation.
% Participants found it crucial to consider this subjectivity since the success of \textbf{Pin of Shame depends on the moral outrage} of the audience and their subsequent pushback against the pinned comment.
For example, Participant P1 is an Azerbaijani living and working as a creator in South Korea and targeting Korean audiences. She noted that her experiences from a different cultural context sometimes made it difficult for her to predict whether certain norm violations in comments would outrage her audiences. Thus, deploying the Pin of Shame was particularly challenging for her, as she had to carefully consider how the Korean audiences might respond.

\begin{quote}
\textit{    ``Of course, like we should do it [comment pinning] smartly. If I think some comment is negative, it's not negative for everyone. So I should be more objective, like, what if I pin these comments, and people will more like opposing with me [sic]. It's not what I want.'' -- P1}
\end{quote}

Many participants observed that they frequently \textbf{pin and unpin} different negative comments, and thereby control which comments to emphasize at any moment, as a way to shape the overall direction and tone of discourse in the comments section. For example, P6 shared her experience of pinning and unpinning negative comments:

\begin{comment}
\textit{    ``I sometimes pin for a day and let people attack this one person. And then, when I'm tired of that person, I'll unpin it, and then pin another.'' -- P7}
\end{comment}

\begin{quote}
    \textit{``The only reason I would replace his comment is, if I found a crazier comment that I can pin.'' -- P6}
\end{quote}

In sum, deciding what constitutes a norm violation and when to unpin comments signals creators’ outsized influence in shaping and enforcing local norms within their comment sections. 
However, the publicly mediated nature of social sanctioning within the Pin of Shame distinguishes it from conventional forms of moderation.

\subsubsection{Leveraging Negative Comments for User Engagement and Content Promotion}
\label{sec:User engagement}

Given the commercial logics of social media platforms, most creators are motivated to build greater viewership of their channel and posted content. Interestingly, we found that our participants considered audience engagement with their content even when addressing negative comments using the pinning feature.
Some participants \textbf{believed that news feed algorithms ``love'' negative comments}, and expected that posts with a contentious comments section would be recommended to more viewers. Given this understanding, they sought to leverage negative comments on their posts %to boost user engagement and algorithmically promote post 
by keeping them visible, anticipating that this would further boost user engagement and promote post visibility.
% Thus, they use the Pin of shame to reflect their strategic expectations regarding content virality.
For example, when a video posted by P6 criticizing men who deny women’s abilities received a lot of traffic, she attributed its popularity to the opposing comments:

\begin{quote}
    \textit{``...These [pinned] comments are part of why this video actually went viral, because people just spent so much time in the comments section that the algorithm was like, oh, this is content that people are watching and engaging with.'' -- P6}
\end{quote}

%\begin{quote}
%\textit{    ``Hateful content gets more engagement than positive content. And so it's good for the algorithm. It's good for attention and clicks and stuff like that.'' - P10}
%\end{quote}

% Thus, when deciding which moderation mechanism to deploy in response to inappropriate comments, participants' perceptions of algorithmic virality leads them to choose the Pin of shame over other measures, such as flagging or comment removal.
% Unlike the latter, comment pinning ensures the negative comment remains publicly visible, fostering greater engagement with their own post and its discussion while also punishing and potentially educating the violator.

%\begin{quote}
%    ``So I feel that when I delete comments or block comments or things like that, it tends to potentially affect my engagement.'' - P14
%\end{quote}
Indeed, some participants deliberately \textbf{pin negative comments that have already sparked engagement} or are likely to generate further discussion, anticipating continued user interaction.
%From here
For example, Participant P13 described that when rude comments (e.g., comments criticizing women's appearance) gain popularity through replies, he strategically pins them to maintain their visibility and sustain engagement in the comments section:

\begin{quote}
\textit{     ``So what I see is the one [comment] that is getting the most likes. I'll pin it and make sure it stays at the top. ... I pinned it, because that's the one that I knew most people would see.'' -- P13}
\end{quote}

%\subsubsection{Respecting Free Speech through Allowing Comment Visibility and Counter-speech}
\subsubsection{Preserving Comment Visibility to Respect Free Speech and Maintain Public Record}
\label{sec:Preseving the visibility}

The Pin of Shame socially sanctions the pinned comments while preserving their visibility, meaning that negative comments are not removed from public view.
This contrasts with common moderation practices, which often rely on visibility reducing measures such as reporting or deleting comments.
Participants understood that visibility reduction measures can sometimes lead to a backfire effect, i.e., they receive criticism from commenters and the public for restricting \textbf{free speech}.
For this reason, many participants appreciated the Pin of Shame for its ability to impose punishment without forcibly removing comments.

\begin{quote}
    \textit{``I never delete comments [so] that people can speak their mind because I like the idea of freedom of speech. You're okay to say anything, because once you take that away from someone, then there are so many other things you can take away.'' -- P16}
\end{quote}

By keeping inappropriate comments visible, many creators saw comment pinning as opening up a space in which the broader audience could respond, challenge, and deliberate over the comment.
They emphasized that allowing pinned commenters to express their perspectives and enabling counter-speech from the broader audience better aligned with their understanding of free-speech principles.
For example, P10 explained that she rarely removes negative comments unless they contain explicit slurs, and prefers to keep them visible because she believes excessive removal can becomes a ``slippery slope'' of over moderation:

\begin{quote}
    \textit{``Sometimes it can be dangerous to start deleting things, because it's like a slippery slope. [With pinning,] other people can see it, and they can have their thoughts on it, and I have my thoughts on it, and that's fine for me.'' -- P10}
\end{quote}

Moreover, some participants highlighted the importance of \textbf{preserving negative comments for their documentation value}.
Keeping such comments visible allowed them to remain as part of the digital record, serving as enduring evidence that could be seen and interpreted by others over time.
Participant P19 mentioned:

\begin{quote}
    \textit{``When I see those negative comments, I think it's such a ridiculous comment. And so I want them to know it's a ridiculous comment in a way that you put that out there as your digital footprint.'' -- P19}
\end{quote}

% \subsection{How Comment Pinning Affordances Shape the Use of the Pin of Shame}
% \subsubsection{Flexible and Easy to Use Moderation Tool}
% \label{sec:Flexible and easy to use}

% Due to its easy-to-use nature, 

%Moreover, given the limitation that Instagram allows \textbf{pinning only up to three comments} at a time, participants who post on that site feel a need to be more deliberate and selective in choosing which comments to pin.

\subsection{Consequences of Deploying the Pin of shame}

% This strategy is often successful---the Pin of shame frequently prompts comment viewers to both post \textbf{counterarguments against the negative commenter} and \textbf{defend the original poster (OP)}, transforming the comments section into a discursive space to confront, challenge, and correct inappropriate behavior.

\subsubsection{Holding Norm Violators Accountable}
\label{sec:Accountable through shaming}

Consistent with participants' expectations, the Pin of Shame often results in collective shaming, i.e., \textbf{other users
post counterarguments against the negative commenter} and \textbf{defend the original poster}, transforming the comments section into a discursive space to confront, challenge, and correct inappropriate behavior.
% criticize or express their disapproval of the pinned comment.
Participants viewed this collective shaming as an \textbf{apt punishment for the pinned commenters}, a \textit{fair} response to their infractions, and an effective way to hold them accountable. For instance, Participant P15 made an analogy between such shaming and legal punishment by emphasizing how both measures make violators face the consequences of their actions:

\begin{quote}
    \textit{``Maybe it's [the Pin of Shame is] like community service when you get caught doing something stupid. It's like you go to the judge, and the judge says `You were wrong. Here's your 25 hours of community service.' Maybe the 25 hours of community service is 25 of my followers calling that person ridiculous for making that statement.'' -- P15}
\end{quote}

In response to such shaming, the pinned commenters sometimes \textbf{delete the pinned comment}  (e.g., in the case of Jayan) or \textbf{apologize} to participants.
However, in other cases, the Pin of Shame incites either no reaction from the pinned commenters or sparks an intense argument between them and the shaming public.
Participants appreciated that using the Pin of Shame still affords them an opportunity to deploy additional moderation measures when the resulting discourse from the Pin of Shame does not yield the expected outcomes.
In the latter cases, participants usually follow up by implementing another sanction, such as reporting or deleting the concerning comment:

\begin{quote}
    \textit{``I think that [the Pin of Shame] doesn't always work, and there are instances where I do delete them, typically if they start insulting other people who are not me. You never know how someone else is handling that, or how that might affect them.'' -- P4}
\end{quote}

\subsubsection{Attaining Emotional Empowerment and Social Support}
\label{Sec:emotional empowerment}
Some participants described initially using flags, the key platform-offered feature to address inappropriate content, to report norm violations. However, this usually
resulted in the platform refusing to take any action on the flagged item.
Participants interpreted such refusals as a dismissal of their emotional and moral judgment, and felt that platform-offered tools like flags fail to address or even acknowledge their harm.
Such experiences motivated them to pursue making sense of their harm by making inappropriate comments on their posts hyper-visible. 
For example, P5 described reporting a comment against her transgender identity only to be told that it was not considered hate speech:

\begin{quote}
    \textit{``The comment that I pinned was `Gas chamber'. I reported that comment, and Instagram said that wasn't hate speech or reportable and so after that, I pinned the comment ... I just can't believe we live in a world where you can say things like that to people.'' -- P5}
\end{quote}

%Many participants felt that using the Pin of Shame reduces the emotional labor required to address negative comments just by oneself. 
Pinning enabled creators to invite public attention and observe how others interpret and react to the pinned comment.
Several participants described perceiving public pushback against pinned comments as a form of emotional validation of their experiences.
Reading through such pushback enabled them to \textbf{make sense of their hurt emotions} by confirming that the harm they experienced reflected a shared norm violation rather than merely a personal sensitivity.
This social support allowed them to feel less isolated in the face of negativity, and leverage their followers' assistance to counteract negative comments.
% Engaging in this practice, followers create a collective atmosphere that makes the creator feel supported and less isolated in the face of negativity. 
% This dynamic shows how the Pin of Shame not only manages negative comments but also fosters solidarity and mutual support within the community. 
Participant P14 shared:

\begin{quote}
    \textit{``It's [Pin of Shame] making it more of a group conversation and kind of also showing me that people have my back, because sometimes I feel like I'm just arguing with people constantly.'' -- P14}
\end{quote}

These experiences of public validation also provided emotional relief to participants. Typically, participants described the Pin of Shame as a satisfying, pleasurable, or enjoyable experience, where the shaming of those who leave negative comments provides catharsis through a general recognition that justice has been served. 
P2 shared:

%\begin{quote}
%    \textit{``Just watch the shit, pin the comment, and watch people respond to that person, it's just kind of a source of amusement sometimes, and makes me feel better when other people who are trying to troll me also get trolled. I don't know. It's probably not like the best strategy. But it makes me feel better sometimes. I won't lie. ... It's very satisfying to see.'' -- P14}
%\end{quote}

\begin{quote}
    \textit{``So to me it's really satisfying, seeing that people will stand up against a negative comment and back me up. It feels good to see that people will not stand for bullying.'' -- P2} 
\end{quote}

Participants also noted that these \textbf{emotional benefits extend beyond themselves.} 
Some of them observed that when negative comments containing misogyny or hate speech are pinned, it explicitly encourages targeted in-groups to unite among themselves, foster a sense of community, and resist online hate.
Participants expected that in such cases, collective actions by their followers may allow them to experience empowerment (P4) and an emotional catharsis (P3) at their success in shaming the norm violators.
Participant P4, who frequently posts content with \#WomenInMaleFields hashtag, explained that pinning misogynistic comments often offers emotional support to her young female STEM followers, as it creates opportunities for them to speak up for one another:

\begin{quote}
     \textit{``I think it's good for young women, because most of the people who follow me are young women who are in STEM or interested in STEM. I think it's good for them to see someone stand up for them.'' -- P4}
\end{quote}

Similarly, P3 described her account as a ``playground'' for her feminist allies, noting that her followers would experience emotional gratification when collectively acting against pinned misogynistic comments:

\begin{quote}
    \textit{``Other users, for example, the women who support me, would definitely press the "like" button [to counter-speeches to the pinned comment] a lot. And I expect that they also feel some kind of pleasure or gratification, similar to what I feel.'' -- P3}

\end{quote}

\begin{comment}

\begin{quote}
    \textit{``I feel like it's a form of entertainment for the people who see it. It's a form of like inspiration, because I do get messages from my followers saying `Yo, bro, keep it up. Keep Moggin,' like.'' -- P16}
\end{quote}
\end{comment}

% Some participants found using the Pin of Shame a means to \textbf{unite their supporters} and consolidate their support.
% For example, 
% Participant P4, whose followers are primarily women, explained that when she pins misogynistic comments, her followers push back against them; she believes that this collective action strengthens her followers' sense of belonging to a community and offers them a source of emotional support.
% % and by pinning the , their followers can observe and collectively push back against it, strengthening their sense of community belonging and providing emotional support to each other.

\subsubsection{Saving the Cognitive Effort to Address Negative Comments}
\label{sec:cognitive effort}

Using the Pin of Shame enables participants to punish or educate negative commenters without direct engagement. It allows them to \textbf{save time and mental exertion} by relying on the involvement of other users, and delegating emotional labor to them.
This contrasts with other coping mechanisms, such as replying to or reporting the negative commenters, which require personal involvement.
Participant P20 highlighted how pinning helps creators manage time constraints and avoid the cognitive burden of responding to numerous comments:

\begin{quote}
    \textit{``%Because when I understand that I cannot do it, 
    Especially when there are like 100 comments, I cannot reply and fight with someone [those commenters], and I have other stuff to do in my life, like commenting on someone. So I would just pin.'' -- P20}
\end{quote}

% Beyond reducing creators' effort, the Pin of Shame also delegates the labor of social sanctioning to the audience, distributing emotional labor across followers through collective participation.
% As P17 described:

% \begin{quote}
%     \textit{``And they're basically doing the work [replying to pinned commenter] for me when people are arguing with each other in my comments like they're doing the work for me. It's like passive income, right? ...  I'm sit back and let everyone else do the work for me.'' -- P17}
% \end{quote}

\begin{comment}
    \begin{quote}
    \textit{``So, I pin the type of comment like `You are old. You look 45' ... People who try to support me and protect me, they will start fighting with that comment as well. So it's kind of like a battle of two people. And I'm just here with my popcorn.'' -- P11}
\end{quote}
\end{comment}

\subsubsection{Engaging in Impression Management}
\label{sec:impression management}
Beyond its immediate use as a moderation tactic, the Pin of Shame often had secondary consequences for how creators managed their public image and interacted with audiences.
Many participants acknowledged that after they pin a negative comment, they often engaged in a playful or lighthearted replies, frequently leaving witty remarks.
This practice often resulted in \textbf{a more relaxed and entertaining atmosphere}, shifting the focus away from the rigid tone of commenter being punished through shaming.
Through such discursive efforts, pinning negative comments creates a space where both the pinned commenter and the public could express their viewpoints more comfortably.
For example, P17, a medical professional, described an instance where she pinned and responded to a comment making unsubstantiated claims about nurses' mistreatment of patients. Her response had the effect of presenting her as avoiding suppression of discussion on this serious topic while making her position on it clear in a light-hearted fashion:

\begin{quote}
    \textit{``When I'm replying back [to pinned comment] and I'm saying like funny things, like, `I'll see you on church on Sunday' ... The way that I respond to them, it shows a lot about my character, and who I am, and people who do support me think it's funny, and they appreciate that.'' -- P17}
\end{quote}

This behavior also reflects participants’ efforts to \textbf{craft an aloof persona} when responding to negative comments.
Rather than engaging in direct confrontation or displaying emotional distress, the Pin of Shame allowed them to employ humor or sarcasm to maintain emotional composure and preserve a positive self-image in public settings.
By reframing negative comments as sources of amusement, creators sought to invert the power dynamic, strategically signaling emotional resilience and control.
For instance, P6 said:
% described how she shifts the tone of negative comments by responding in a playful and humorous way, deflecting pinned negative comments:

\begin{quote}
    \textit{``I'm never serious with them. I always just like joke around. When someone says something like, `Oh, you're not pretty enough to have this opinion', [I respond] `Stop! You're gonna make me sad' with a sad face [emoji] like I'll just joke back with them because I feel like they're trying to hurt me, or they're trying to put their stuff out in the world, and I'm like, I don't care.'' -- P6}
\end{quote}

\subsubsection{Experiencing Emotional Stress}
\label{sec:Emotional stress}
While the Pin of Shame often provides creators and their supporters with positive emotional experiences and strategic benefits, it can also result in unintended negative outcomes.
Some participants shared that when they noticed that the pinned comment continued to receive greater support than their own or their supporters' counter-arguments, it caused them \textbf{emotional stress and discomfort}.
% Participants reflected on the number of likes the pinned comment received, often comparing it to the likes on their own or their supporters' comments in the same comments section.
This was especially true for P19, a new content creator with limited experience managing comments. When an inappropriate comment she had pinned continued to receive support, P19 felt particularly stressed:

\begin{quote}
    \textit{``I pinned it, and I was really scared because at first you can see like this [pinned] comment has about, almost 4,000 likes, and so it's quite a lot actually for a comment. I don't think it's very nice. ... I was kind of worried that people were like really agreeing with this guy [the pinned commenter] because I was like, `Wow, there's too many people that are not nice in this world.' '' -- P19}
\end{quote}

Some participants lamented experiences when the pinned commenter became a target of excessive insults and mockery.
Participants also expressed concerns about instances when the 
argument initiated by the Pin of Shame attracts more haters commenting to support the pinned comment, or when the pinned commenter actively engages by doubling down on their original response.
In such cases, the comments section often devolves into \textbf{mutual insults and hostility}, ultimately amplifying the negativity and disrupting the otherwise constructive dialogue in that section.
% , participants sometimes perceived that the Pin of Shame fails to produce a corrective or educational effect on the rule violator.
% When the violator refused to concede or acknowledge their fault, the conversation frequently devolved into personal attack, with some participants observing that the violator became the target of excessive insults and mockery.
% Some participants reported that they had not anticipated such an escalation and lamented the lack of constructive dialogue, noting that the comments section often \textbf{devolved into mutual insults and hostility, ultimately amplifying the negativity}.
Participant P4 illustrated how replies to one of her pinned comments escalated into a hostile discourse.
She read aloud the chain of replies, originating from the pinned comment in response to a video expressing her perspective on discriminatory attitudes toward lesbian women:

\begin{quote}
    \textit{``Now they're arguing with each other. [Narrating the replies to their pinned comment---] `It's always the men who look like you, insane ratio.' `You look like someone who listens to Juice WRLD and cries.'  I think they go to the person's profile and [find] something to insult.'' --  P4}
\end{quote}

When the negative comment remains fixed at the top and continues to fuel flame wars, some participants expressed concerns about the \textbf{emotional burden of being exposed to negativity} every time they open the comments section.
%Given this risk, some participants highlighted the importance of emotional resilience when using the Pin of Shame. 
Experienced content creators, in particular, emphasized the importance of emotional resilience when using the Pin of Shame. 
For example, P7, a creator with over 10 years of experience creating social justice content, described how she manages this emotional burden while addressing negative comments.

\begin{quote}
    \textit{ ``You have to have a strong backbone to let people, hundreds and thousands of people say negative things to you sometimes at once. You have to like, get past it and then laugh at it, and it takes a long time to get to that point.'' -- P7}
\end{quote}

Moreover, a few participants worried that pinning negative comments may overshadow their original content by placing it in a negative light, and cause the entire post to be perceived as unpleasant.

\begin{quote}
    \textit{``At first I was, I kind of do feel bad. That I make someone [referring to the pinned commenter] in the spotlight and it's like kind of negative, or I was scared like it makes this [post] into a silly video, a negative video.'' -- P19}
\end{quote}

% Given such experiences, some participants considered the Pin of Shame a double-edged sword. 
% While it can lead to positive outcomes, \textbf{it can also bring unexpected negativity}. As such, participants could not always anticipate or control other users' reactions. 
% This leads participants to worry about further harassment and the risk of spreading negativity to others when deploying the Pin of Shame.

% \begin{quote}
%     \textit{``But in some cases I don't think it really is very constructive to be pinning negative comments. ... It can be a slippery slope. It can go one of two ways, where it can be kind of constructive like that, or it can be really, not positive at all. Not constructive.'' - P10}
% \end{quote}
\section{Discussion}

Our analysis of the motivations and consequences of using the Pin of Shame offers key insights into content creators' comment moderation needs, safety strategies, and impression management concerns. These insights offer valuable considerations for HCI scholars and practitioners in developing approaches to address online harms, including through the design of novel content moderation tools. 
% We elaborate these arguments below.

\subsection{Benefits and Risks of Deploying Public Shaming to Punish and Educate Norm Violators}
Our analysis shows that the Pin of Shame practice aligns with previously described instances of public shaming---both online and offline~\citep{ronson2015so,braithwaite1993shame}---in its operating logic. 
As ~\citet{trottier2018coming} observes, shaming practices rely on \textit{visibility}, which holds targets accountable to broader publics through collective criticism.
Our findings similarly show that the Pin of Shame leverages hyper-visibility to render inappropriate commenters accountable to other social actors by attracting the latter's counterarguments (Section~\ref{sec:Public attention}). 
By doing so, this practice allows the community to engage in norm education~\cite{sherif1936psychology,kiesler2012regulating}, reduce the cognitive and emotional labor of moderation on creators~\citep{jhaver2023personalizing,dosono2019moderation,li2022all}, and exert social pressure on pinned commenters into changing their behavior while supporting their free speech.
Through discussing real-world examples with the involved creators of how shaming can function positively, this study provides evidence supporting its potential benefits as documented in prior literature~\cite{scheff2014ubiquity,trottier2018coming,hou2017socioeconomic,kasabova2021shame,direk2020politics,corry2021screenshot,braithwaite1989crime}, particularly when it operates as a crowd-sourced form of governance.
% This balances the dominant narrative in prior research, which has primarily emphasized the negative rhetoric and consequences of shaming~\cite{billingham2020enforcing,lewis2021we,marwick2021morally,trottier2017digital,johnston1996vigilantism,basak2019online,kim2022does}, while also contributing to the body of literature that explores its positive aspects~\cite{scheff2014ubiquity,trottier2018coming,hou2017socioeconomic,kasabova2021shame,direk2020politics,corry2021screenshot,braithwaite1989crime}.

% The motivation and consequences of the Pin of Shame fundamentally rely on online shaming. 
% It aims to persuade rule violators to change their behavior~\cite{shea2024discursive, muir2021portrayal}, and internalize the normative behavior education~\cite{trottier2018coming}, through collective criticism and counterargument from the public~\cite{basak2019online,nussbaum2019shaming} (Section~\ref{sec:Public attention}). 
% This participatory form of shaming~\cite{braithwaite1989crime} also reduces moderation burdens for creators (Section~\ref{sec:social validation}).

% We found that deploying  
% Prior research has shown that responding to harmful comments by reporting or removing them often denies creators the ability to heal from the emotional impact of the harm~\citep{xiao2023addressing,salehi2023sustained,thomas2022s}.
% In contrast, 
The content creators we interviewed for this study valued the \textbf{empowering} property of the Pin of Shame: instead of silent comment removals or account bans (Section~\ref{Sec:emotional empowerment}), it allowed them and their followers to publicly voice their stance and speak out against inappropriate behaviors (Section~\ref{sec:Accountable through shaming}). 
Drawing on Rappaport’s conceptualization of empowerment as gaining greater influence over issues that matter to individuals and communities~\cite{rappaport1985power}, we understand empowerment here as creators’ and their audiences’ increased capacity to define, voice, and respond to harmful behavior. 
This form of empowerment was particularly meaningful when platforms declined to moderate comments that creators experienced as harmful.
From creators' perspective, such strategic online shaming can be especially meaningful for creator groups (and their audience members) who are disproportionately targeted by abuse and harassment (e.g., based on their gender or ethnicity). It provides creators a way to feel emotionally validated rather than to silently endure hurtful comments~\cite{murumaa2021misogynist,im2022women}.
This practice also allows creators the freedom to define norm violations in ways not delimited by platform's code of conduct (Section \ref{sec:what constitutes norm}).
Thus, the Pin of Shame offers a unique set of affordances and benefits within the toolkit of moderation strategies available to creators.

By analyzing this practice, we also contribute to prior discussions on the decentralization of governance power and responsibility across multiple levels of platform governance~\citep{jhaver2023decentralizing,zhang2024form}. 
When this power is concentrated at the top-level, users often report dissatisfaction with moderation decisions~\citep{duffy2023platform,myers2018censored,shahid2023decolonizing}.
In contrast, in our case study, content creators serve as the middle level units of governance and exert control over regulating comments in their purview.
Notably, they do so in this case by appropriating the comment pinning feature in ways that run counter to its normative use as prescribed by the platform.
This represents an instance of creators leading the charge of moderation via seizing visibility control, a function usually limited to and associated with platforms' centralized recommendation apparatus.
This \textbf{shift in the governance authority} presents novel tradeoffs: on the positive side, middle level units hold a significant degree of autonomy from above, i.e., the platform admins level, in enacting visibility-based local norm regulations and bypassing impenetrable platform policies. However, this unilateral authority to shape visibility and encourage retribution-based counterspeech also creates novel accountability challenges. 

% as it not only enacts sanctioning but also empowers users to voice their concerns and collectively educate norm violators.
%Our research findings shed light on the constructive utility of shaming, particularly its role in fostering collective resistance against hate speech and discrimination targeting disproportionately affected groups in online spaces.

Indeed, collective shaming by the public is inherently uncalibrated, carrying the potential for conversations derailing or escalating into \textbf{retributive harassment}~\cite{blackwell2018online,marwick2021morally,lewis2021we,klonick2015re}. 
Shaming also focuses attention on a target rather than on their infraction---it assumes that the pinned comment captures the whole of the commenter's character and denies them the benefit of the doubt. 
Our participants reported that the Pin of Shame can devolve into doxxing or personal insults, demonstrating the risk that such practices can transform into harassment.
These unintended negative consequences of the Pin of Shame can cause \textbf{emotional stress} for content creators (Section~\ref{sec:Emotional stress}) as well as pinned commenters.

\subsection{Shame as a Justice Practice in Platform Governance}
Creators adopt the Pin of Shame to make norm violating actions visible, thereby eliciting public shaming 
(Section~\ref{sec:Public attention}).
This public shaming symbolically challenges violators' moral standing within the community by publicly inscribing their fault. In meting out this punitive yet socially mediated sanction in response to the fault, the Pin of Shame practice can be consistent with principles of retributive justice ~\citep{govier2011forgiveness,wenzel2008retributive}.
% Thus, the Pin of Shame constitutes a form of punitive social sanctioning grounded in retributive justice.

Existing governance models primarily respond to online hate and harassment~\cite{thomas2022s} by sanctioning perpetrators through content removal or access restriction~\cite{cai2024content,schoenebeck2021drawing}. 
%These practices reflect criminal justice models that focus on punishment~\cite{im2022women}.
Prior HCI and CSCW research has shown that such punitive approaches
%---which are easily bypassed by creating new posts or accounts---
often fail to acknowledge the experiences of targeted users or hold perpetrators accountable for their actions~\cite{xiao2023addressing,jhaver2018online,salehi2023sustained}.
In contrast, we found that the Pin of Shame allows content creators to obtain emotional validation, and perceive \textbf{social support} through audience participation (Section~\ref{sec:Public attention}).
Therefore, when enacted through participatory and socially mediated designs, as in the Pin of Shame, retributive punishment can afford opportunities for sensemaking, validation, and emotional accountability that conventional punitive moderation measures often fail to provide.

Consequently, our findings point to the need to reconsider how punitive moderation is enacted and experienced.
Specifically, when creators can validate their experiences of harm and observe how norms are publicly interpreted and negotiated through audience engagement (Section~\ref{sec:Public attention}), shaming practices may function for them as a more emotionally satisfying punitive response.

This aligns with prior CSCW research identifying situations in which shaming responses to online harms are perceived as beneficial~\citep{schoenebeck2021drawing,sultana2021unmochon}. 
For example, \citet{im2022women} show that users consider revealing perpetrators’ real names and photographs—a practice relying on public shaming to invoke vigilante justice—as a desirable response to online harassment.
These insights suggest that platforms and creators should reconsider not just whether punitive logics are appropriate, but how such logics are implemented and experienced.
Similar to restorative justice approaches that emphasize the emotional experiences of those harmed~\citep{xiao2023addressing,salehi2023sustained,schoenebeck2021drawing}, our findings indicate that retributive responses can also incorporate emotional validation and sensemaking when socially mediated.
At the same time, participants’ careful considerations of when to deploy the Pin of Shame underscore that shaming-based punitive approaches are compelling only under specific conditions and require creators to be sensitive to their audience's values and norms. 
%Rather, they point to the risks of operationalizing retributive logics abstracting them into \textbf{one-size-fits-all enforcement mechanism}~\citep{jiang2021understanding,rabaan2021daughters,sultana2021unmochon,im2022women} that strip punishment of visibility, communication, and social context. 

\subsection{Moderation Design Should Attend to Creators' Impression Management Goals During Norm Enforcement}
In addition to providing emotional support, we found that the Pin of Shame operates as a tool for impression management~\citep{goffman2023presentation}.
It allows content creators to incorporate entertainment into the moderation process, such as by adding playful remarks (Section~\ref{sec:impression management}). 
This lighthearted approach aims to craft a more relaxed atmosphere and a favorable self-presentation to the public (Section~\ref{sec:impression management}).
Thus, the Pin of Shame serves the performative needs~\cite{li2019live,pellicone2017game} of creators confronting negativity by allowing them to redirect it into positivity. 
It enables them to frame their responses to stress in ways that can be perceived as resilience or equanimity within their public persona.

While prior research has examined impression management through various dimensions of online self-presentation---such as signaling political affiliation through `Likes'~\citep{marder2016like} or showcasing lists of interests (books, music, movies)~\citep{liu2007social} on social media---our study reveals that even harm addressing and moderation practices can become intertwined with creators' impression management goals within the platform economy. 
However, such performative considerations do not necessarily align with the traditional goals of content moderation, such as reducing harm or ensuring public safety. 
These tensions highlight the need to design creator moderation with a firm understanding of a broader array of creator concerns, and accounting for how audience perceptions and performative incentives may shape creators' moderation practices.

We especially encourage designers to consider how public, audience-facing forms of social sanctioning introduce new dynamics---in displays of expression, strength, and agency---that complicate the relationship between moderation, impression management, and creator/public safety.
These insights also raise crucial further questions about \textbf{how creators' impression management goals factor into} their adoption of \textbf{other forms of creator-led moderation} actions (e.g., replying, deleting) when addressing online harms.

\subsection{Recommender Systems Should Enact and Communicate the Use of Emotion-Sensitive Engagement Metrics to Shape Visibility}
Our participants often shared a folk theory that negative comments are likely to trigger algorithmic amplification of their posted content.
Regardless of the extent to which this folk theory~\cite{gelman2011concepts,french2017s} 
aligns with actual news feeds outputs, its frequent occurrence in our interview data suggests that it motivates creators to deploy public shaming in an effort to exploit the recommendation algorithm, with the hope that it would promote their content to a broader audience (Section \ref{sec:User engagement}).
% Our findings are meaningful in that they engage with existing research on relationship between negativity and content viewership—such as negative contents attracts more user engagement~\cite{crockett2017moral}, thus tends to reach broader audiences due to higher rates of dissemination and sharing~\cite{brady2017emotion}—by suggesting that similar dynamics may be perceived to operate even in micro-contexts like the comments section.
% However, this folk theory underscores the need for empirical research into whether negativity in user responses, not just in primary content, actually influences recommendation algorithms.
Indeed, we found that some creators pin negative comments, even when doing so takes an emotional toll, such as repeated exposure to negativity or argumentative threads sparked by the pinned comment (Section~\ref{sec:Emotional stress}).
Such actions underscore a limitation in current algorithmic design: recommendation systems often give content creators the perception that their underlying algorithms prioritize engagement metrics without considering whether the interactions are constructive or hostile.

Prior CSCW literature has documented that algorithmic content feeds can induce emotional distress for users with histories of eating disorders~\cite{gak2022distressing}, those exposed to content about their ex-partners~\cite{pinter2019never}, and those targeted with ads that tell them what is ``wrong'' with them~\cite{schoenebeck2025algorithmic}. 
Extending this research, our findings reveal another instance of algorithmically induced content-based harm.
We found that in some cases, \textbf{creators self-inflict exposure to negative comments}.
Despite being aware of the possible emotional costs (although the exact extent of such costs may be harder to predict beforehand), creators weight these harms against potential enhancements in visibility or audience engagement when enacting the Pin of Shame, effectuating their folk-theory that the recommendation algorithm incentivizes controversy.

Importantly, creators show different levels of discomfort they are willing to endure. In our data, experienced creators described developing emotional resilience that enabled them to better manage sustained exposure to negativity, whereas less experienced creators were more likely to feel overwhelmed and to withdraw by unpinning comments or closing discussions (Section \ref{sec:Emotional stress}). Thus, the tradeoff between attempts at reducing harm exposure and influencing algorithms for public attention is mediated by creators’ experience levels and emotional resilience.
% Thus, creator experience shapes how algorithmic incentives are negotiated, or resisted.

Based on these findings, we emphasize the need for \textbf{emotion-sensitive algorithmic design} for content recommendations, one that not only captures engagement volume but also accounts for the emotional tone of user interactions. 
Fostering creators' understanding and trust in such emotionally aware systems would assuage their ``algorithmic anxiety''~\citep{jhaver2018airbnb} and
disenthrall them from enduring negative comments just to please news feed algorithms.
This may shift platform dynamics toward more positive and sustainable participation, and better support the long-term emotional well-being of content creators.

%\subsection{Visibility as a Tool for Participatory Moderation}

\subsection{Moderation Tools Should Seek to Incorporate Transparency Regarding the Sanctioning Processes}
The Pin of Shame represents a unique form of moderation that maintains the visibility of inappropriate comments, rather than removing them outright (Section~\ref{sec:Public attention}). 
By preserving the original comment, creators can avoid the potential backlash regarding free speech suppression that is often associated with actions that trigger removal (such as reporting, deleting, or muting)~\cite{myers2018censored,zhang2023cleaning} while simultaneously maintaining the comment’s documentation value (Section~\ref{sec:Preseving the visibility}).
The sanctioning occurs via the ensuing discussion publicly signaling the inappropriateness of the pinned comment, similar to how soft moderation approaches clarify norm violations in the labeled content~\citep{zannettou2021won,morrow2022emerging}.
% ---while also enabling content creators to impose further sanctions on the comment if they deem it necessary (Section~\ref{sec:Preseving the visibility}).

By making the entire sanctioning process visible to the public, the Pin of Shame increases the transparency of the norm enforcement process. Our participants appreciated that bystanders can observe the context of the original inappropriate statement, witness how social sanctioning is carried out, and examine the outcomes, such as whether it leads to an apology (Section~\ref{sec:Accountable through shaming}).
% Since transparency in the entire moderation process, including the context behind sanctioning decisions, is of key value for users~\cite{jhaver2019did, shim2024incorporating, zhang2023cleaning, myers2018censored}, our study suggests the value of the Pin of Shame for its transparency as a moderation tool.
% 
This is in line with prior research by ~\citet{sasse2023breaking}, who demonstrated that bystanders prefer context-preserving sanctions through an online experiment; specifically, they concluded that silent deletion of sexist content is judged as less fair than counterspeech against visible sexist content.

\subsection{Norm Negotiation by Content Creators and the Publics}
% Since individual users’ moral leanings toward a pinned comment may vary, the consequences of deploying the Pin of Shame do not always meet the expectations of content creators (Section~\ref{sec:what constitutes norm}). 
% % This variation surfaces how people hold differing views on what constitutes a norm violation, especially when disagreement arises over the appropriateness of the comment sanctioning.
% % 
% These moments of divergence suggest that the comment section can serve not only as a site for moral judgment, but also as a \textbf{contested space} where norms are collectively negotiated. 
% Given that perceptions of harm and what to do about them are shaped by cultural contexts~\cite{jiang2021understanding} and individual experiences~\cite{jhaver2017view}, the significance of the Pin of Shame lies more in its function as an arena where differing interpretations of norm violations become visible.
% This exposes a demand for spaces that enable a public deliberation of which behaviors deserve to be sanctioned.

%For example, when creators notice that a pinned comment has generated controversy over its appropriateness, they can provide context beneath the pinned comment, explaining why they deem it problematic. 
%Such descriptive clarifications may help align diverse perspectives and contribute to a more transparent norm-setting process.

Unlike conventional moderation systems, where users have limited options---such as appeals---to express disagreement with enforcement decisions~\cite{vaccaro2020end,haimson2021disproportionate,clune2024content}, the Pin of Shame offers a more open and visible arena for users to engage in a nuanced collective debate, which creators appreciate.
% Users’ desire for spaces for norm negotiation is not limited to user-led harm-addressing practices.
% Prior research on social media flagging mechanisms similarly shows that users want to reference others' opinion when evaluating norm violations, for example by discussing the appropriateness of reporting specific content in forums or by referencing prior flagged cases and their outcomes~\cite{shim2024incorporating}. 
%Across both contexts, these demands center on the visibility and referentiality of norm judgments, enabling users to share, compare, and reflect on how norms are interpreted and enforced.
%More broadly, we found that the Pin of Shame functions as a catalyst for making visible these differing norm constructions. 
% Especially in highly context-reliant cases, this format encourages participatory norm construction~\cite{billingham2020enforcing,kasabova2021shame}.
% Such interactions may foster mutual understanding and lay the groundwork for a more consensual and inclusive approach to defining what is acceptable in online spaces. 
% 
% Taken together, these insights suggest that 
This surfaces the limits of unilateral norm enforcement and point toward the need for more dialogic approaches to norm setting and enforcement.
Platforms and users alike should work to foster a culture in which norms are co-constructed through open dialogue informed by diverse criteria, lived experiences, and cultural backgrounds~\cite{direk2020politics,billingham2020enforcing}. 
Such an approach could support a healthier online environment by acknowledging the persistence of moral disagreement and by making conflicts around norms more visible and open to scrutiny.

Currently, implementation of the Pin of Shame inherently concentrates \textbf{disproportionate power in the hands of content creators}, as only they can initiate the practice by highlighting a particular comment.
Further, the affordances of the comment section---which allow creators to arbitrarily pin or unpin any user comment and remove any comment---may be exploited to selectively shape discourse and steer public interpretation toward the creator's perspective.

While many participants described their use of the Pin of Shame as justice-oriented punishment, we cannot assume creators to be uniformly benevolent actors. 
Given that even well-intended or self-protective actions can escalate into harassment~\citep{freeman2025have}, and that retributive harassment is often perceived as justified or deserved by observers~\citep{blackwell2018online}, we argue that the Pin of Shame is particularly susceptible to such escalation. 
% By enabling creators to publicly foreground specific comments through unilateral control over visibility, pinning can concentrate attention and scrutiny on an individual commenter, creating conditions under which justice-oriented actions may transform into targeted harassment.
Indeed, prior CSCW research highlights that platform affordances that facilitate visibility and coordination can foster conditions under which justice-oriented actions may transform into targeted harassment~\citep{blackwell2018online, marwick2021morally}. 
This raises the question of how to distribute moderation power and responsibilities between platforms and creators such that creators’ authority remains both legitimate and consistent with platform-level norms~\citep{jhaver2023decentralizing}.
% Such amplification affordances may further facilitate collective pile-ons directed toward individuals~\citep{kim2022does}.

We argue that lightweight design interventions that enhance creator accountability and bound the reach of pinned comments could be valuable in this regard.
For example, when a pinned comment generates controversy regarding its appropriateness, creators could be nudged to provide clarification beneath the pinned comment, explaining why they consider it problematic. 
%Second, platforms could restrict the sharing of pinned comments beyond their original context, which could help prevent implicit calls to action. 
Additionally, introducing temporal throttling on replies to pinned comments may disrupt rapid collective pile-ons and constrain escalation.
% We argue that such design affordances can help constrain persistence, coordination, and escalation, and support a more transparent and reflective process of collective norm negotiation.

\subsection{Limitations and Future Work}\label{sec:limitations}
We interviewed only the content creators who deployed the Pin of Shame.
While this sampling strategy allowed us to closely examine comment pinning as a moderation practice from creators' perspective, due to social desirability bias, our participants may have emphasized the perceived benefits and value of this practice, and perhaps under-reported the negative ways (e.g., enacting harassment) in which they used comment pinning.
Further, our purposive sampling approach likely underrepresents less visible, smaller-scale, private, or less algorithmically amplified creators who may have engaged in this practice. 
Collecting and critically analyzing a larger corpus of Pin of Shame deployments would complement our research by providing further valuable insights.

In addition, 
% our analysis focuses primarily on platform–creator dynamics and does not fully capture how follower–creator relationships shape the outcomes of the Pin of Shame. 
it would be useful to examine the perspectives of pinned commenters to understand how they perceive the experience of being publicly shamed, and how it shapes their attitudes about future posting on social media.
Large-scale, longitudinal analyses of pinned commenters' behaviors can also reveal the effects of Pin of Shame on norm violators' posting patterns.

\section{Conclusion}
This study examines the Pin of Shame, a retributive-justice-oriented creator-led moderation strategy that relies on audience engagement to enforce channel norms. Drawing on interviews with 20 content creators who deployed this strategy, we demonstrate how it reconfigures punitive moderation logics through a communicative and transparent approach.
Building upon our findings, we contribute theoretical insights regarding the use of shame as a justice practice for content moderation.
We also offer design suggestions to help prevent such shaming interventions from escalating into harassment.
We highlight how creators' impression management goals serve a key role in shaping their moderation choices. 
Finally, we show how creator-led shaming practices are shaped by platform affordances and power asymmetries. This underscores the need for multi-level governance that supports norm negotiation while holding creators accountable for their actions.

\begin{acks}
Awaiting paper acceptance.
\end{acks}

\bibliographystyle{ACM-Reference-Format}
\bibliography{references}

\appendix

\section{Participant Sample} \label{sec:appendix-participants}
Table \ref{tab:demographic} shows the demographic characteristics and channel features of the interview participants. 

\begin{table}[h]
    \setlength{\arrayrulewidth}{0.25mm}
    \centering
    \small
    \renewcommand{\arraystretch}{1.1} % spacing
    \begin{tabularx}{\textwidth}{ccc>{\raggedright\arraybackslash}Xc>{\raggedright\arraybackslash}Xc>{\raggedright\arraybackslash}X}
        \hline
        \textbf{No.} & \textbf{Age} & \textbf{Gender} & \textbf{Occupation} & \textbf{Country} & \textbf{Platform} & \textbf{Size} & \textbf{Topic} \\
        \hline
        P1  & 29  & Female & Student & South Korea & Instagram, Tiktok, Youtube & 5k-10k & Beauty, Travel \\
        P2  & 27  & Female & Executive Assistant & US & Instagram, Tiktok & 20k-50k & Social issue \\
        P3  & 26  & Female & Musician & South Korea & Instagram & 1k-5k & Feminism \\
        P4  & 31  & Female & Postdoctoral Researcher & US & Instagram & 50k-100k & Lifestyle \\
        P5  & 31  & Female & IT Industry Professional & US & Instagram, Tiktok & 5k-10k & Entertainment \\
        P6  & 27  & Female & Consultant & US & Instagram, Tiktok, Youtube & 500-1k & Media \\
        P7  & 57  & Female & Performer & US & Instagram & 20k-50k & Social Justice \\
        P8  & 21  & Female & Student & US & Instagram, Tiktok & 5k-10k & Beauty, Fashion \\
        P9  & 26  & Female & Model & US & Instagram & 5k-10k & Mental health, Profession \\
        P10 & 21  & Non-binary & Social Worker & US & Instagram & 1k-5k & Entertainment \\
        P11 & 37  & Female & Content Creator & US & Instagram, Tiktok & >100k & Entertainment \\
        P12 & 19  & Female & Student & Hong Kong & Instagram & 20k-50k & Lifestyle \\
        P13 & 21  & Male & Musician & US & Instagram & 1k-5k & Wellness \\
        P14 & 38 & Female & Stylist & US & Instagram, Tiktok & 50k-100k & Hobbies \\
        P15 & 35  & Female & Vice President of Marketing & US & Instagram & 10k-20k & Beauty, Wellness \\
        P16 & 23  & Male & Sales Representative & US & Instagram, Tiktok & 10k-20k & Fitness, Travel \\
        P17 & 32  & Female & Medical Professional & US & Instagram & 20k-50k & Profession \\
        P18 & 36  & Female & Medical Professional & US & Instagram, Tiktok & 50k-100k & Lifestyle \\
        P19 & 21  & Female & Student & Hong Kong & Instagram & 1k-5k & Lifestyle \\
        P20 & 22  & Female & Unemployed & Serbia & Instagram & 50k-100k & Travel \\
        \hline
    \end{tabularx}
    \caption{Demographic information of interview participants. In the table, `Platform' refers to all the platforms where participants share the content they create (with the first listed platform being their primary channel, which is the one they focus on most and upload content to first). `Size' refers to the number of followers on their primary channel. `Topic' denotes the main focus or theme of their channel.}
    \label{tab:demographic}
\end{table}

\end{document}